\documentclass[preprint,12pt]{elsarticle}

\usepackage{lineno,hyperref}
\usepackage{amssymb}
\usepackage{amsfonts}
\usepackage{lipsum}
\usepackage[normalem]{ulem}
\usepackage{amsmath}
\usepackage{xcolor}

\enlargethispage{\baselineskip}

\hypersetup{
     colorlinks   = true,
     citecolor    = blue
}

\journal{Physics of the Dark Universe}

\begin{document}

\begin{frontmatter}



\title{The future search for low-frequency axions and new physics with the FLASH resonant cavity experiment\\ at Frascati National Laboratories\\ 
}
 
\author[inst1]{David Alesini}

\author[inst1]{Danilo Babusci}
\author[bel1]{Paolo Beltrame}
\author[inst1]{Fabio Bossi}
\author[inst1]{Paolo Ciambrone}

\author[inst1]{Alessandro D'Elia}
\ead{alessandro.delia@lnf.infn.it}

\author[inst1]{Daniele Di Gioacchino}
\author[inst1]{Giampiero Di Pirro}

\author[maxplanck]{Babette D\"obrich} 
\author[fbk]{Paolo Falferi}

\author[inst1]{Claudio Gatti}
\author[Ba,CAPA]{Maurizio  Giannotti}
\author[inst1]{Paola  Gianotti}

\author[pi1]{Gianluca Lamanna}
\author[inst1]{Carlo Ligi}
\author[inst1]{Giovanni Maccarrone}
\author[inst1]{Giovanni Mazzitelli}
\author[unibari,ibari]{Alessandro Mirizzi}
\author[squid]{Michael Mueck}

\author[inst1,b]{Enrico Nardi}
\author[enea]{Federico Nguyen}
\author[inst1]{Alessio Rettaroli}
\author[unicam,inst1]{Javad Rezvani}

\author[pi2]{Francesco Enrico Teofilo}
\author[inst1]{Simone Tocci}
\author[inst1]{Sandro Tomassini}

\author[c,d]{Luca Visinelli}
\author[c,d]{Michael Zantedeschi}

\affiliation[inst1]{organization={INFN, Laboratori Nazionali di Frascati},
            addressline={via Enrico Fermi 54}, 
            city={Roma},
            postcode={00044}, 
            country={Italy}}
\affiliation[bel1]{organization={University of Liverpool Department of Physics},
             addressline={Oxford St, L69 7ZE},
             city={Liverpool},
             postcode={},
             country={England}}

\affiliation[maxplanck]{organization={Max-Planck-Institut  für Physik (Werner-Heisenberg-Institut)},
             addressline={Föhringer Ring 6},
             city={München},
             postcode={80805},
             country={Germany}}
\affiliation[fbk]{organization={Fondazione Bruno Kessler},
             addressline={Via Sommarive, I-38123},
             city={Povo, Trento},
             postcode={},
             country={Italy}}

\affiliation[Ba]{organization={Department of Chemistry and Physics, Barry University},
            addressline={11300 NE 2nd Ave.}, 
            city={Miami},
            postcode={33161}, 
            country={USA}}

\affiliation[CAPA]{organization={Centro de Astropartículas y Física de Altas Energías (CAPA), Universidad de Zaragoza},
            city={Zaragoza},
            postcode={50009}, 
            country={Spain}}

\affiliation[pi1]{organization={INFN and University of Pisa},
            addressline={Largo Pontecorvo 3}, 
            city={Pisa},
            postcode={56127}, 
            country={Italy}}
            
\affiliation[unibari]{organization={Dipartimento di Fisica "Michelangelo Merlin"},
            addressline={Via Amendola 173}, 
            city={Bari},
            postcode={70126}, 
            country={Italy}} 
            
\affiliation[ibari]{organization={INFN sezione di Bari},
            addressline={Via Orabona 4}, 
            city={Bari},
            postcode={70126}, 
            country={Italy}} 
            
\affiliation[squid]{organization={ez SQUID},
             addressline={Herborner Strasse 9},
             city={Sinn},
             postcode={35764},
             country={Germany}}
\address[b]{Laboratory of High Energy and 
         Computational Physics, HEPC-NICPB, \\
        R{\"a}vala 10, 10143, Tallinn, Estonia}
            
\affiliation[enea]{organization={ENEA Centro Ricerche Frascati},
            addressline={Via E. Fermi 45}, 
            city={Frascati},
            postcode={I-00044 }, 
            country={Italy}}     

\affiliation[unicam]{organization={Physics Division, School of Science and Technology, Università di Camerino},
             addressline={Via Madonna delle Carceri 9},
             city={Camerino},
             postcode={62032},
             country={Italy}}

\affiliation[pi2]{organization={University of Pisa},
            addressline={Largo Pontecorvo 3}, 
            city={Pisa},
            postcode={56127}, 
            country={Italy}}
\affiliation[c]{organization={Tsung-Dao Lee Institute (TDLI)},
            addressline={520 Shengrong Road}, 
            city={Shanghai},
            postcode={201210}, 
            country={China}}

\affiliation[d]{organization={School of Physics and Astronomy, Shanghai Jiao Tong University},
            addressline={800 Dongchuan Road}, 
            city={Shanghai},
            postcode={200240}, 
            country={China}}

\begin{abstract}
We present a proposal for a new experiment, the FINUDA magnet for Light Axion SearcH (FLASH), a large resonant-cavity haloscope in a high static magnetic field which is planned to probe new physics in the form of dark matter (DM) axions, scalar fields, chameleons, hidden photons, as well as high frequency gravitational waves (GWs). Concerning the QCD axion, FLASH will search for these particles as the DM in the mass range $(0.49-1.49)\,\mu{\rm eV}$, thus filling the mass gap between the ranges covered by other planned searches. A dedicated Microstrip SQUID operating at ultra-cryogenic temperatures will amplify the signal. The frequency range accessible overlaps with the Very High Frequency (VHF) range of the radio wave spectrum and allows for a search in GWs in the frequency range $(100-300)\,$MHz. The experiment will make use of the cryogenic plant and magnet of the FINUDA experiment at INFN Frascati National Laboratories near Rome (Italy); the operations needed to restore the functionalities of the apparatus are currently underway. We present the setup of the experiment and the sensitivity forecasts for the detection of axions, scalar fields, chameleons, hidden photons, and GWs.
\end{abstract}


\end{frontmatter}


\section{Introduction}
\label{sec:sample1}

The evidences for the existence of Dark Matter (DM) is overwhelming, with observations spanning from the rotation of galaxies~\cite{Rubin:1970zza}, the gravitational lensing of light~\cite{Trimble:1987ee}, and the large-scale structure of the Universe~\cite{Efstathiou:1992sy, Planck:2018nkj} which all point to the presence of a vast amount of invisible matter. Despite these considerations, a non-gravitational probe of the cosmic DM is yet to be obtained, due to its feeble interaction with the Standard Model (SM) degrees of freedom which makes its detection very challenging.

A possible DM candidate is the QCD axion~\cite{Weinberg:1977ma, Wilczek:1977pj}, a light pseudoscalar particle arising within the solution to the strong CP puzzle proposed by Peccei and Quinn (PQ)~\cite{Peccei:1977ur, Peccei:1977hh}. The puzzle emerges as a naturalness problem in the following way. The QCD sector possess a continuum of vacua characterised by a CP-violating angle $\theta$~\cite{Callan:1976je,Jackiw:1976pf}. However, the physical measurable quantity is $\overline{\theta}=\theta + \arg{\det{ M}}$, with $M$ being the mass matrix of quarks. 
The above quantity induces an electric dipole moment in the neutron~\cite{Baluni:1978rf,Crewther:1979pi}, and is severely constrained $\overline{\theta} \lesssim 10^{-10}$~\cite{Baker:2006ts}.

CP-violating contributions to the electric dipole moment of the neutron are also induced by the SM electroweak interaction. However, these effects are significantly smaller~\cite{Ellis:1976fn,Shabalin:1979gh,Ellis:1978hq} than the current phenomenological reach. Nevertheless, it is somewhat puzzling that two separate unrelated contributions produce such a minuscule quantity.
 
The axion makes the angle $\overline{\theta}$ dynamical and relaxes the vacuum structure to the CP-invariant sector. The mechanism relies on a $U(1)$ global symmetry broken by the QCD anomaly. Any other source of explicit breaking would reintroduce CP violation, therefore posing a challenge for the mechanism which, in order to explain a tiny number, requires the exactness of its global symmetry. This is even more evident when considering gravity. In fact, any global $U(1)$ symmetry is expected to be explicitly broken by quantum gravity effects~\cite{Kallosh:1995hi}. This leads to the so-called axion-quality problem~\cite{Kamionkowski:1992mf, Ghigna:1992iv, Barr:1992qq}. Several mechanisms have been proposed to protect the symmetry from this sort of explicit breaking, see e.g.\ Refs.~\cite{2018tasi.confE...4H, DiLuzio:2020wdo}.\footnote{Recently it has also been argued that consistency with quantum gravity forces the axion mechanism to be exact, see~\cite{Dvali:2018dce,Dvali:2022fdv}.}

Although the original QCD axion model proposed in Refs.~\cite{Weinberg:1977ma, Wilczek:1977pj} has been long excluded, other models in which the axion feebly couples to the SM degrees of freedom have been proposed, such as the Kim-Shifman-Vainshtein-Zakharov (KSVZ) model~\cite{Kim:1979if, Shifman:1979if} and the Dine-Fischler-Srednicki-Zhitnitsky (DFSZ) model~\cite{Zhitnitsky:1980tq, Dine:1981rt}. These benchmark models belong to a larger class which is referred to as the ``invisible'' axion models. A rich experimental program will probe the existence of the QCD axion in the next decade~\cite{Adams:2022pbo,Baryakhtar:2022hbu,Antel:2023hkf}. Among the experiments, ADMX~\cite{ADMX:2001dbg, ADMX:2001nej, ADMX:2010prl, ADMX:2018ogs, ADMX:2018gho, ADMX:2019uok, ADMX:2021nhd, ADMX:sidecar}, HAYSTAC~\cite{AlKenany:2016trt, Brubaker:2016ktl, HAYSTAC:2018rwy, HAYSTAC:2020kwv, HAYSTAC:2023cam}, ORGAN~\cite{McAllister:2017lkb, McAllister:2020twv,ORGAN:scienceAdv}, BabyIAXO~\cite{Ahyoune:2023gfw}, the facilities at IBS-CAPP~\cite{Lee:2020cfj, Jeong:2020cwz, CAPP:2020utb, Lee:2022mnc, Kim:2022hmg, Yi:2022fmn}, CAST-CAPP~\cite{Adair:2022rtw}, RADES~\cite{ArguedasCuendis:2019swy, Melcon:2018dba, AlvarezMelcon:2020vee, CAST:2020rlf}, QUAX~\cite{Barbieri:2016vwg, Crescini:2018gwh, Alesini:2019ajt, Crescini:2018qrz, QUAX:2020adt, Alesini:2020vny, Alesini:2022lnp, DiVora:2023dzs}, DMRadio/ABRACADABRA~\cite{Kahn:2016aff, Foster:2017hbq, DMRadio:2022pkf}, CADEx~\cite{Aja:2022csb}, GrAHal~\cite{Grenet:2021vbb} and TASEH~\cite{TASEH:2022vvu} will use a haloscope, i.e.\ a detector composed of a resonant cavity immersed in a strong magnetic field~\cite{Sikivie:1983ip, Sikivie:1985yu, Krauss:1985ub, Hagmann:1989hu}.

Along with these proposed or ongoing haloscope activities is the FINUDA (FIsica NUcleare at DA$\Phi$NE) magnet for Light Axion SearcH (FLASH) experiment at INFN Frascati National Laboratories (INFN-LNF) near Rome (Italy), which will probe the existence of cosmic axions of masses around $10^{-6}\,$eV. This window is currently unexplored and lies in between the mass range that is actively scanned in present and near-future searches by   ADMX \cite{ADMX:2021nhd}, BabyIAXO~\cite{Ahyoune:2023gfw} and DMRadio~\cite{DMRadio:2022pkf} collaborations. In this view, FLASH will close the mass gap in the range of the $\mu$eV where the QCD axion is expected to provide the DM budget.

FLASH will also be able to constrain cosmological scenarios in which exotic particles other than the QCD axion play a role. One such example includes axion-like particles (ALPs)~\cite{DiLuzio:2016sbl, DiLuzio:2017pfr, DiLuzio:2017ogq}, see also the various reviews on the subject~\cite{Raffelt:1995ym, Raffelt:2006rj, Sikivie:2006ni, Kim:2008hd, Wantz:2009it, Kawasaki:2013ae, Marsh:2015xka, Irastorza:2018dyq}. Alternative scenarios predict that the DM is composed of scalar rather than pseudoscalar particles. In this work, we assess the reach of the FLASH instrumentation when exploring models of the dilaton~\cite{Cho:2007cy, Choi:2011fy} and the chameleon~\cite{Khoury:2003aq, Brax:2004qh}, two scalar fields. Another avenue that motivates the FLASH experiment is related to hidden photon (HP) DM~\cite{Redondo:2010dp, Arias:2012az, Graham:2015rva}, see also Ref.~\cite{Caputo:2021eaa}. In fact, haloscope experiments can efficiently constrain the mixing parameter describing the coupling between the SM photon and the cosmic HP in the viable phenomenological window.

Finally, it has been recently realized that resonant cavities could also serve the purpose of detecting gravitational waves (GWs) signals within the band ranges falling in the MHz-GHz region~\cite{Aggarwal:2020olq}. 
High-frequency GWs in the kHz range from coalescent compact objects or making up a stochastic background have been sought for in past experiments Explorer~\cite{Astone:1993ur} at CERN and Nautilus~\cite{Astone:1997gi} at INFN-LNF. Here, we show that the FLASH experiment can probe the GW bandwidth $\sim (100-300)$\,MHz. As discussed below, exploring this frequency range may provide insights on the abundance of primordial black holes (PBHs).

The paper is organized as follows. In Sec.~\ref{ch1:sec_sum-halo} we discuss the design and the forecast reach of the FLASH haloscope. In Sec.~\ref{ch1:sec_theory} we review the cosmology of the axion and we report the potential results for the search of QCD axions with FLASH. The forecast results for other models such as an axion-like particle, the hidden photon, chameleons, and GW signals are discussed in Sec.~\ref{sec:othermodels}. The details of the design and tuning for the radio frequency cavity are given in Sec.~\ref{sec:RF}, while the cryogenics is discussed in Sec.~\ref{sec:cryogenics}. The methods developed for the acquisition and analysis of the data are reported in Sec.~\ref{sec:signalacquisition} and Sec.~\ref{sec:dataanalysis}, respectively. Finally, our conclusions are drawn in Sec.~\ref{sec:conclusions}. We work in natural units $\hbar = c = k_B = \varepsilon_0=  1$ unless otherwise specified.

\section{Summary of the FLASH haloscope forecast reach}
\label{ch1:sec_sum-halo}

The FLASH experiment is expected to investigate the existence of the axion in the mass region between $m_a=0.49{\rm\,\mu}$eV and $m_a=1.49{\rm\,\mu}$eV by using resonant cavity techniques. Such a haloscope apparatus foresees a copper resonant cavity with an inner volume of approximately 4.15\,m$^3$ in the first stage of the search. The material and the properties of the cavity walls affect the cavity's efficiency in storing energy, which is quantified by the {\it unloaded} quality factor $Q_0$. The {\it loaded} quality factor $Q_L=Q_0/(1+\beta)$ accounts for the additional coupling between the cavity and the receiver by a quantity $\beta$. Coordinates and orientation of the detector are shown in Tab.~\ref{tab:orientation}.

\begin{table}[!ht]
  \begin{center}
    \caption{Position and magnetic-field direction  of FLASH experiment}
    \label{tab:orientation}
    \begin{tabular}{c|c}
    \hline
    Latitude &  $41^{\circ}$ $49^{\prime}$ $26^{\prime\prime}$ \\
      Longitude & $12^{\circ}$ $40^{\prime }$ $13^{\prime\prime}$ \\
      Elevation & 100~m\\ 
      B field direction & East$-$northeast  \\
      \hline\hline
    \end{tabular}
  \end{center}
\end{table}

\begin{figure}[!ht]
  \begin{center}
    \includegraphics[totalheight=6cm]{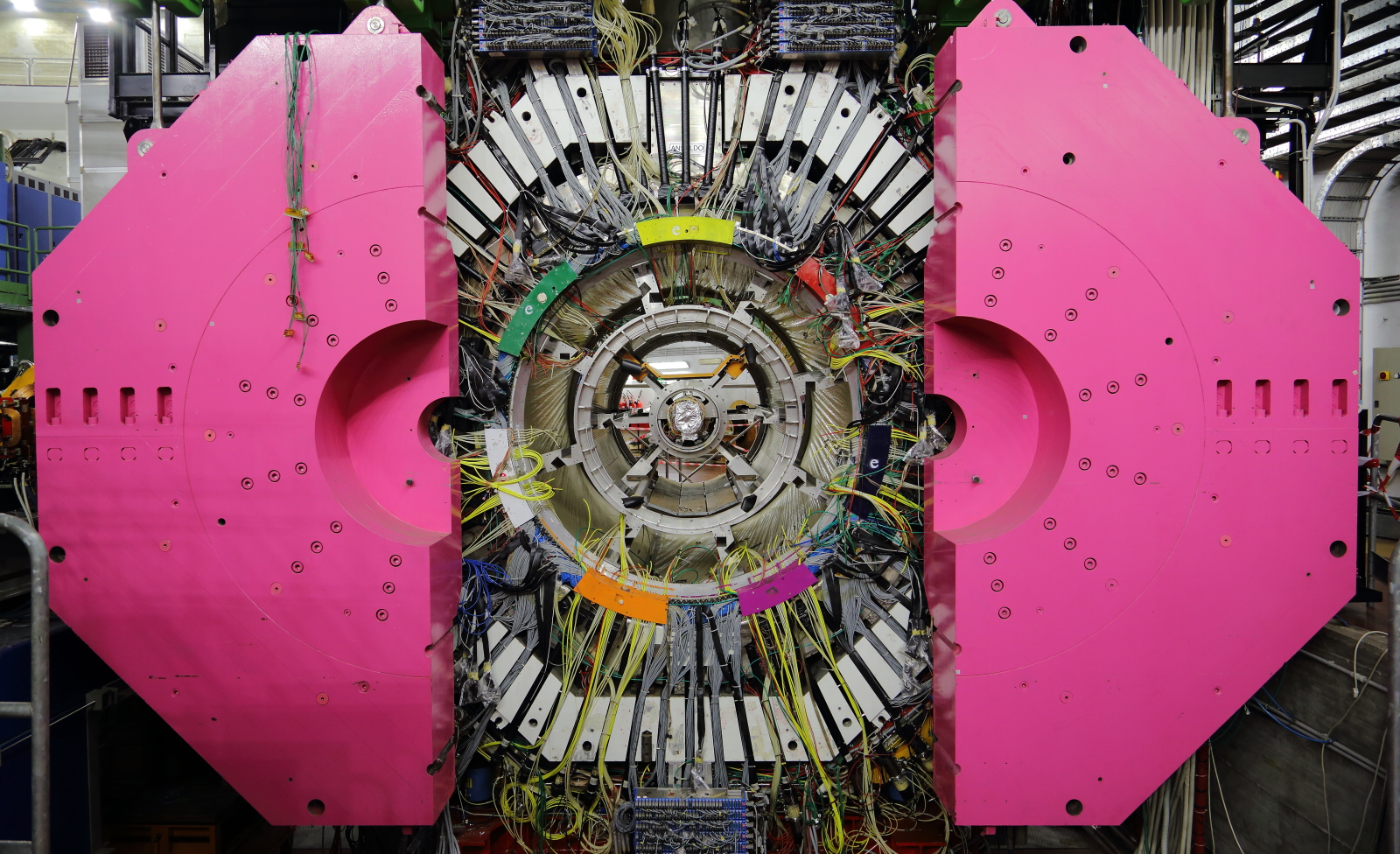}
    \caption{The FINUDA apparatus.}
    \label{fig:kloemag}
  \end{center}
\end{figure}

The axion converts in the presence of the strong magnetic field provided by the FINUDA magnet~\cite{bertani1999finuda, Modena:1997tz}, an iron shielded solenoid coil of 1.4\,m in radius and 2.2\,m in length, made from an aluminium-stabilised niobium titanium superconductor. The magnet can provide a homogeneous axial field of strength up to $B_0$ = 1.1\,T. The field homogeinity, $\delta B_Z/B_0<5\%$ and $\delta B_X/B_0<1\%$, is provided by the two iron end-caps that close the magnet yoke.
See Fig.~\ref{fig:kloemag} for additional details.

The expected power $P_{\rm sig}$ for a QCD axion converting inside a cavity resonating at the frequency $\nu_c$ can be as low as $10^{-22}\,$W, see Eq.~\eqref{eq:power} below. 
Therefore the cavity has to be cooled to cryogenic temperatures to efficiently detect the signal. 
For this, ultra-low noise cryogenic-amplifiers are needed for the first stage amplification, with the cryostat apparatus cooled down to 4.5\,K being inserted inside the bore of the FINUDA magnet.

According to the Dicke radiometer equation~\cite{Dicke:1946glx}, the signal to noise ratio (SNR) is
\begin{equation}
    \label{eq:snr}
    {\rm SNR} = \frac{P_{\rm sig}}{ T_{\rm sys}}\sqrt{\frac{\tau}{\Delta\nu_I}}\,,
\end{equation}
where $T_{\rm sys}$ is the combination of the amplifier and the thermal noises, $\tau$ is the integration time and $\Delta\nu_I$ is the intrinsic signal bandwidth. For galactic axions, the latter corresponds to $\Delta\nu_I = \nu_c/Q_a$, where the astrophysical factor $Q_a \simeq 10^{6}$ accounts for the DM velocity dispersion. The integration time for each value of the scanning resonant frequency and for a given SNR can be determined from inverting Eq.~\eqref{eq:snr},
\begin{equation}
    \label{eq:integrationtime}
    \tau = ({\rm SNR})^2 \frac{\nu_c}{Q_a} \left(\frac{ T_{\rm sys}}{P_{\rm sig}}\right)^2\,. 
\end{equation}
If the loaded quality factor of the cavity is smaller than $Q_a$, the number of signal bandwidths that can be scanned simultaneously by a cavity is $\sim Q_a/Q_L$. The scan rate is obtained as the ratio of the frequency step $\Delta \nu = \nu_c/Q_L$ and the scan time in Eq.~\eqref{eq:integrationtime} to obtain (see also Refs.~\cite{Stern:2015kzo, AlKenany:2016trt})
\begin{equation}
    \label{eq:rate}
    \frac{{\rm d}\nu}{{\rm d}t} \approx \frac{Q_a}{Q_L}\,\left(\frac{P_{\rm sig}}{{\rm SNR}\, T_{\rm sys} }\right)^2\,.
\end{equation}
As discussed in Ref.~\cite{Stern:2015kzo}, the scan rate refers to the minimum target for the axion coupling and thus provides a measure of the sensitivity of the experiment.

Once the signal has originated in the cavity it has to be amplified and picked up efficiently. A Microstrip SQUID Amplifier (MSA)~\cite{Muck:1999ts, Muck_2010} operating at 300\,mK is an optimal solution, in terms of low noise, frequency band and gain, for the first stage of signal amplification. In the 4\,K region a cryogenic heterojunction field-effect transistor (HFET) amplifier is employed. A summary of amplification steps and the equivalent temperature noise is shown in Table~\ref{tab:noise}, with a discussion of the amplification stage being given in Sec.~\ref{sec:amplifiers}.
\begin{table}[!ht]
  \begin{center}
    \caption{Summary of amplification steps and equivalent noise temperature.}
    \label{tab:noise}
    \begin{tabular}{c|c|c|c}
      Device &  Gain & Noise Temperature & operating temperature \\ \hline
      MSA & 20 dB & 0.4\,K & 0.3\,K \\
      HFET & 15 dB & 5\,K & 4.5\,K \\
      Secondary Amplification & 60 dB & 150\,K & 300\,K \\
      \hline\hline
    \end{tabular}
  \end{center}
\end{table}

\section{Models of the QCD axion in the range of FLASH}
\label{ch1:sec_theory}

\subsection{Basic concepts}

In the following three sections, we discuss the physical scenarios accessible to FLASH. We begin with the QCD axion, arguably one of the most motivated and studied new-physics candidate. At energies below the PQ and the electroweak symmetry breaking scales, the dynamics of the axion field $a$ is described by an effective Lagrangian
\begin{equation}
    \label{eq:lagrangian}
    \mathcal{L} = \frac{1}{2}(\partial^\mu a)(\partial_\mu a) + \frac{\alpha_s}{8\pi}\frac{a}{f_a}\tilde G^{\mu\nu}G_{\mu\nu} + \frac{1}{4}g^0_{a\gamma\gamma} a \tilde F^{\mu\nu} F_{\mu\nu} + \frac{1}{2f_a}(\partial_\mu a) j_{a,0}^\mu\,,
\end{equation}
where $\alpha_s$ is the strong force coupling strength, $f_a$ is the QCD axion decay constant, 
$F^{\mu\nu}$, $G^{\mu\nu}$ are the electromagnetic (EM) and gluon field strengths respectively, and a tilde indicates the dual of the field strengths. The coupling of the axion to the photons is described by the coupling constant
\begin{equation}
    \label{eq:bare_coupling}
    g^0_{a\gamma\gamma} = \frac{\alpha_{\rm EM}}{2\pi f_a} \frac{E}{N}\,,
\end{equation}
where $\alpha_{\rm EM}$ is the fine-structure constant and $E/N$ is the ratio of the EM and the color anomalies. Different QCD axion models predict different values of the ratio $E/N$. The last term in Eq.~\eqref{eq:lagrangian} describes any model-dependent coupling of the axion with the SM fermions, with fermionic current $j_{a,0}^\mu$.

The QCD axion mass originates from non-perturbative effects during the QCD phase transition~\cite{Weinberg:1977ma}. After a rotation of the quark fields, the interaction of the axion field with the chiral condensate implies a potential for the QCD axion of the form~\cite{DiVecchia:1980yfw, GrillidiCortona:2015jxo}
\begin{equation}
	\label{Vqcd}
    V(a) = -m_\pi^2f_\pi^2\,\sqrt{1-\frac{4m_um_d}{(m_u+m_d)^2}\,\sin^2\left(\frac{a}{2f_a}\right)}\,,
\end{equation}
where $m_\pi$ and $f_\pi$ are the mass and the decay constant of the pion and $m_u$, $m_d$ are the masses of the up and down quarks, respectively. The axion mass at zero temperature from the mixing with the neutral pion is then~\cite{Weinberg:1977ma}
\begin{equation}
	\label{eq:m0}
	m_a = \frac{\sqrt{m_u m_d}}{m_u+m_d}\,\frac{m_\pi f_\pi}{f_a} \equiv \frac{\Lambda^2}{f_a},
\end{equation}
where $\Lambda \approx 75.5\,$MeV is an energy scale related to the QCD phase transition.

In the new quark field basis, the axion-photon coupling is redefined as
\begin{eqnarray}
    g_{a\gamma\gamma} &=& \frac{\alpha_{\rm EM}}{2\pi f_a}\left( \frac{E}{N} - \frac{2}{3} \frac{4 m_d + m_u}{m_d + m_u}\right) \equiv \frac{\alpha_{\rm EM}}{\pi f_a}\,g_\gamma\,,\\
    g_\gamma &\equiv& \frac{1}{2}\left( \frac{E}{N} - \frac{2}{3} \frac{4 m_d + m_u}{m_d + m_u}\right)\,,
\end{eqnarray}
where the model dependent parameter is $g_{\gamma} = -0.97$ $(g_{\gamma} = 0.36)$ in the KSVZ (DFSZ) axion model~\cite{Kim:1979if, Shifman:1979if, Dine:1981rt, Zhitnitsky:1980tq}. The properties of the QCD axion today are determined by the behavior of its potential with temperature, which controls both the dynamics of the axion field in the early Universe as well as the production of topological defects~\cite{Vilenkin:1982ks, Sikivie:1982qv}.

At temperatures higher than that of the QCD phase transition, the effects of instantons become severely suppressed. This reflects onto the value of the axion mass, whose dependence on the temperature is fixed in terms of the QCD topological susceptibility. The axion mass decreases quickly with an increasing value of the temperature of the plasma above the confinement temperature $T_C$~\cite{Gross:1980br}. Generally, the QCD axion mass is expressed in terms of the QCD topological susceptibility $\chi(T)$ as
\begin{equation}
	\label{eq:QCDaxion_mass}
    m_a^2(T) = \frac{m_a^2}{\Lambda^4}\,\chi(T)\,,
\end{equation}
where $\chi(T)$ is normalised such that $\chi(T=0) = \Lambda^4$. Estimating the temperature dependence of the topological susceptibility is one of the goals of QCD lattice simulations that capture the dynamics of the quark-gluon plasma around $T_C$~\cite{Bonati:2015vqz, Borsanyi:2015cka, Borsanyi:2016ksw, Petreczky:2016vrs}.

\subsection{Detecting cosmic axions with FLASH}

The axion-photon coupling sparks hope to detect cosmic axions that make up the DM in the Galaxy by means of resonant cavities on Earth. FLASH is expected to operate within the EM resonant frequency of $(117 - 360)\,$MHz, corresponding to an axion in the mass range $m_a\simeq (0.49 - 1.49)\,\mu$eV. The search is divided into a low-frequency (LF) region within $(117 - 206)\,$MHz and a high-frequency (HF) region within $(206 - 360)\,$MHz, with the corresponding mass ranges within $(0.49 - 0.85)\,\mu$eV and $(0.85 - 1.49)\,\mu$eV, respectively. 
When the resonant frequency of the cavity $\nu_c$ is tuned to the corresponding axion mass $m_a/(2\pi)$, the expected power deposited by DM axions is~\cite{AlKenany:2016trt, Brubaker:2016ktl}
\begin{equation}
	\label{eq:power}
	P_{\rm sig} = \left( g_{\gamma}^2\frac{\alpha_{\rm EM}^2}{\pi^2}\frac{\rho_a}{\Lambda^4} \right) \times
	\left( \omega_c Q_L\,\frac{\beta}{1+\beta} B_0^2 V C_{mnl} \right)\,.
\end{equation}
Here, $\rho_a$ is the local axion density, and the second set of parentheses contains the magnetic field strength $B_0$, the cavity volume $V$, the angular frequency $\omega_c=2\pi\nu_c$, and a geometrical factor $C_{mnl} \simeq O(1)$ that depends on the cavity mode.

\begin{figure}[!ht]
  \begin{center}
    \includegraphics[width=0.8\linewidth]{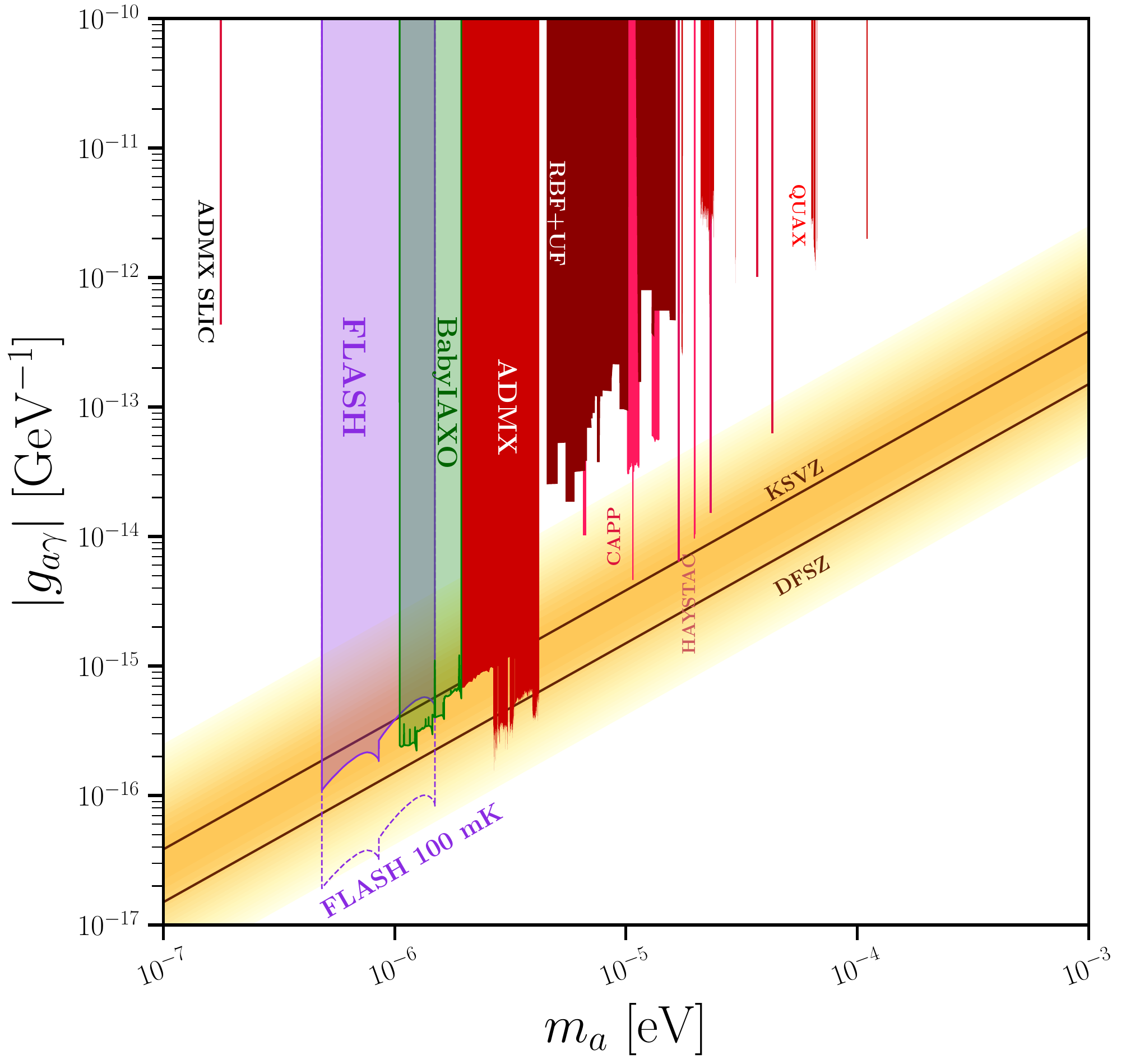}
    \caption{The FLASH discovery potential (90\% confidence level or c.l.) compared to existing experimental limits. The brown lines with yellow error-band show the theoretical predictions for the KSVZ and DFSZ axions~\cite{Kim:1979if, Shifman:1979if, Dine:1981rt, Zhitnitsky:1980tq}. The forecast reach of FLASH is compared with experimental limits from other haloscopes~\cite{ADMX:2001dbg, ADMX:2001nej, AlKenany:2016trt, Brubaker:2016ktl, McAllister:2017lkb, McAllister:2020twv, DePanfilis:1987dk, Wuensch:1989sa, Hagmann:1990tj} as well as a projection from \cite{Ahyoune:2023gfw} labled `babyIAXO' in green, which is expected to be realized somewhat later than FLASH. Image realized with~\cite{AxionLimits}. }
    \label{fig:sensitivity_zoom}
  \end{center}
\end{figure}

The discovery potential in the coupling-mass plane calculated through Eqs.~(\ref{eq:snr}) and (\ref{eq:power}) is shown in Fig.~\ref{fig:sensitivity_zoom}. The FLASH sensitivity can reach the band predicted for QCD axions down to the KSVZ model line. 
It is assumed here that axions make up the totality of the dark matter, with the local DM energy density fixed to the reference value $\rho_{\rm DM} = 0.45{\rm\,GeV\,cm^{-3}}$~\cite{Peng:2000hd,Buch2019,BENITO2021100826}.
We considered an integration time of 5 minutes for a single measurement with the large cavity and 10 minutes for measurements for the small cavity. In Table~\ref{tab:sensitivity} we take as an example the search in the larger volume setup for the frequency $\nu_c = 150\,$MHz, corresponding to a KSVZ axion of mass $m_a = 0.62\,\mu$eV and coupling $g_{a\gamma\gamma}^{\rm KSVZ} = 2.45 \times 10^{-16}{\rm\,GeV^{-1}}$. The parameters used in Eq.~\eqref{eq:power} are given in the table, with the scan rate obtained from Eq.~\eqref{eq:rate}. According to the estimated number of frequency steps shown in Table~\ref{tab:tuning} the total integrated time will be about 2 years. We also show the potential reach with a cryogenic system upgraded to operate at a lower temperature $T_{\rm sys} = 100\,$mK  pointing to the possibility of probing the DSFZ-axion model.
\begin{table}[!ht]
    \def\arraystretch{1.2}
    \begin{center}
    \caption{The FLASH discovery potential for KSVZ axions. The choice of the parameters corresponds to the search at the peak frequency $\nu_c = 150\,$MHz in the large volume cavity and for a period $\tau=5\,$min. The coupling $\beta = 2$ is chosen to optimize the scan rate~\cite{AlKenany:2016trt, Brubaker:2016ktl}.}
    \label{tab:sensitivity}
    \vspace*{0.5cm}
    \begin{tabular}{c|c}
      Parameter & Value \\\hline
      $\nu_c$\,[MHz] & 150\\
      $m_a$\,[$\mu$eV] & 0.62\\
      $g_{a\gamma\gamma}^{\rm KSVZ}$\, [GeV$^{-1}$] & $2.45\times10^{-16}$  \\
      $Q_L$ & $1.4\times 10^5$\\
      $C_{010}$ & $0.53$\\
      $B_{\rm max}$\, [T]  & 1.1 \\
      $\beta$ & 2 \\
      $\tau$\, [min] & 5\\
      $T_{\rm sys}$\, [K] & 4.9 \\
      $P_{\rm sig}$\, [W] & $0.9\times10^{-22}$\\
      Scan rate\, [Hz\,s$^{-1}$] & 8 \\
      \hline
      $m_a$\,[$\mu$eV] & 0.49 - 1.49\\
      $g_{a\gamma\gamma}$\, 90\% c.l.\ \, [GeV$^{-1}$] & $\left(1.25-6.06\right)\times10^{-16}$  \\
      \hline\hline
    \end{tabular}
    \end{center}
\end{table}

\subsection{Production of cold axions in the early universe}
\label{sec:production}

Cold axions are produced in the early universe through various mechanisms, which include the vacuum realignment mechanism (VRM)~\cite{Abbott:1982af, Dine:1982ah, Preskill:1982cy} and the decay of topological defects. Both these mechanisms have to be studied jointly, nominally by means of numerical simulations.

The present energy density stored in the coherent oscillations of the field can be obtained by solving the equation of motion for the axion field in an expanding universe,
\begin{equation}
	\label{eq:equation_motion}
	\ddot a - \nabla^2 a + 3H\dot a + \frac{{\rm d}V(a)}{{\rm d} a} = 0\,,
\end{equation}
where a dot means a derivative with respect to cosmic time $t$.

At very high temperatures, the axion potential can be safely neglected, so that the axion field is frozen on super-horizon scales. The axion potential becomes a relevant term in Eq.~\eqref{eq:equation_motion} when the Universe has sufficiently cooled off so that the Hubble rate $H(T)$ at temperature $T$ is of the same order of the axion mass,
\begin{equation}
	\label{eq:condition_oscillations}
	3H(T_{\rm osc}) \approx m_a(T_{\rm osc})\,,
\end{equation}
where $T_{\rm osc}$ is the temperature at which the coherent oscillations in the axion field begin. If the transition takes place in the standard cosmological scenario, the number of axions in a comoving volume is preserved to present time, so that the present abundance of axions is expressed as
\begin{equation}
    \label{eq:abundance}
    \Omega_a = \frac{m_a(0)}{m_a(T_{\rm osc})} \frac{\rho_a(T_{\rm osc})}{\rho_{\rm crit}} \frac{s(T_0)}{s(T_{\rm osc})}\,.
\end{equation}
Here, $T_0$ is the present temperature of the CMB photons, $s(T)$ is the entropy density at temperature $T$, and $\rho_{\rm crit} = 3H_0^2M_{\rm Pl}^2$ is the present critical density in terms of the Hubble constant $H_0$ and the reduced Planck mass $M_{\rm Pl}$. 
The axion energy density at the onset of field oscillations $\rho_a(T_{\rm osc})$ is obtained upon solving Eq.~\eqref{eq:equation_motion}.
Thus, it depends on the choice of the potential and of the initial conditions for the axion field $a_i$ which are set at the time of the PQ phase transition. 
We parametrize this by introducing the initial axion angle $\theta_i \equiv a_i / f_a$, whose variance in the distribution across the observable Universe $\langle\theta_i^2\rangle$ reveals information about the state of the early Universe when $T \simeq f_a$.

\subsubsection{Pre-inflation axions}

We first discuss the scenario in which the PQ symmetry breaking happens before or during inflation and it is never restored afterwards. In this scenario, topological defects are inflated away and do not contribute to the QCD axion energy density. One single patch within which the initial axion angle $\theta_i$ has a homogeneous value inflates outside of the scale of the observable Universe, so that the present energy density in axions is found by solving Eq.~\eqref{eq:equation_motion} on super-horizon scales to give (see e.g.\ Ref.~\cite{Visinelli:2009zm})
\begin{equation}
	\label{eq:CDM_axion_mass_ScenarioI}
    m_a \approx 5\,\mu{\rm eV}\,\theta_i^{12/7}\,.
\end{equation}

\subsubsection{Post-inflation axions}

If the PQ symmetry breaking happens after the end of inflation, topological defects such as axionic strings and domain walls form, with properties that greatly depend on the specific model for the QCD axion considered. In this scenario, which is generally referred to as {\em post-inflationary scenario}, $\langle\theta_i^2\rangle$ is the average of the initial VRM angle squared over the circle, assuming that  $\theta_i$ is drawn from a uniform distribution. For a quadratic potential one has $\langle\theta_i^2\rangle= \pi^2/3$, however, in the periodic potential that defines the QCD axion, this result is modified due to the presence of non-harmonic terms~\cite{Turner:1985si}.

The expression in Eq.~\eqref{eq:CDM_axion_mass_ScenarioI} with $\theta_i = \pi/\sqrt{3}$ yields to the na\"ive estimate for the QCD axion mass in this scenario $m_a \approx 20\,\mu{\rm eV}$, which is about one order of magnitude above the reach of FLASH. Including the sub-horizon dynamics and the contributions from topological defects generally leads to a higher value of the axion mass, in the range $m_a = [20-500]\,\mu$eV, according to recent simulations~\cite{Klaer:2017qhr, Klaer:2017ond, Gorghetto:2018myk, Vaquero:2018tib, Buschmann:2019icd, Gorghetto:2020qws, Buschmann:2021sdq, OHare:2021zrq, Hoof:2021jft, Eggemeier:2022hqa}, pushing the window further away from the FLASH sensitivity.

Nevertheless, various uncertainties in the standard cosmological model and the particle content beyond the SM can greatly affect the above picture and modify the expected window of DM axion masses. In this sense, an axion discovery by FLASH would provide insights on the particle content of the early universe.

In the following we briefly discuss possible mechanisms that lower the expected mass window of the QCD axion to the range accessible to FLASH, leaving further details to the literature cited.

\subsection{Scenarios of axion cosmology within the reach of FLASH}
\label{section:III}

\subsubsection{Initial misalignment angle}

In the standard cosmological scenario, the QCD axion is the DM particle with the mass expressed in Eq.~\eqref{eq:CDM_axion_mass_ScenarioI}, when the PQ symmetry is spontaneously broken during inflation. In this scenario, the range of mass sensitivity in FLASH corresponds to the range $|\theta_i| \in (0.2 - 0.4)$. For a uniform probability of $\theta_i \in [-\pi,\pi]$, the probability that $\theta_i$ is drawn with the desired value is about 6\%. Details for the probability distribution derived from the Fokker-Planck equation can be found in Refs.~\cite{Graham:2018jyp, Takahashi:2018tdu} and lead to a somewhat lower probability.

Recently, Ref.~\cite{Hoof:2018ieb} used a Bayesian analysis technique, based on the code GAMBIT~\cite{GAMBIT:2017yxo} and its module DarkBit~\cite{GAMBITDarkMatterWorkgroup:2017fax}, to present a global fit to explore the parameter space of the QCD axion (both the KSVZ and the DFSZ models). In particular, it is considered the scenario in which the Peccei-Quinn symmetry breaks during a period of inflation, while taking into account results from various observations and experiments in the likelihood including the light-shining-through-wall experiments, helioscopes, cavity searches, distortions of gamma-ray spectra, supernovae, horizontal branch stars and the hint from the cooling of white dwarfs. The marginalised posterior distribution obtained in Ref.~\cite{GAMBIT:2017yxo} when demanding that the totality of DM is in axions gives the range $0.12\,\mu{\rm eV} \leq m_a \leq 0.15{\rm \,meV}$ at the 95\% equal-tailed confidence interval. We stress that a portion of the range inferred by this analysis is well within the reach of the FLASH experiment.

Small initial values of $\theta_i$ might also occur naturally, i.e.\ without any fine tuning, in low-scale inflation models in which inflation lasts sufficiently long~\cite{Graham:2018jyp, Takahashi:2018tdu}. 
If $H_I \lesssim \Lambda_{\rm QCD}$ (see e.g. Ref.~\cite{Dvali:1995ce}), the axion acquires a mass already during inflation, the $\theta_i$-distribution flows towards the CP conserving minimum and, for a long duration of the inflation period, stabilises around sufficiently small $\theta_i$ values. As a result the QCD axion can naturally give the DM abundance for axion masses well below the classical window, down to $10^{-12}\,$eV~\cite{Graham:2018jyp}.

\subsubsection{Entropy generation}

If a new species is present in the early universe and if it decays into thermalized products prior to Big Bang Nucleosynthesis (BBN), the entropy density in a comoving volume is not conserved. If a relevant amount of entropy is generated after the axions are produced, for example by the decay of a massive scalar field, the axion density in Eq.~\eqref{eq:abundance} would be diluted by a factor $\Delta$~\cite{Dine:1982ah, Steinhardt:1983ia, Lazarides:1987zf, Lazarides:1990xp,Kawasaki:1995vt, Visinelli:2009zm, Visinelli:2009kt, Visinelli:2017imh}, which would in turn lower the value for the DM axion mass. When $\rho_A^{\rm tot} \to \rho_A^{\rm tot} / \Delta$, the axion mass is
\begin{equation}
	\label{eq:axionmass_diluted}
	m_a \approx \frac{\left(5 - 50\right)\,\mu{\rm eV}}{\Delta^{7/6}}.
\end{equation}
The range explored by FLASH is reached if the contribution from the dilution factor is of order $\Delta \approx \left(10 - 100\right)$. The value of the quantity $\Delta$ depends on the details of the modified cosmological model and, ultimately, on the reheating temperature. A detailed derivation has been given in Refs.~\cite{Visinelli:2009kt, Visinelli:2017imh, Visinelli:2018wza, Draper:2018tmh, Nelson:2018via, Ramberg:2019dgi, Blinov:2019rhb, Ramberg:2020oct, Mazde:2022sdx}.

\subsubsection{Modifying the relation between the axion mass and decay constant}

A different approach to modify the relation between the temperature and the mass of the axion in Eq.~\eqref{eq:QCDaxion_mass} is through the addition of particle content, without changing the cosmological evolution. 
This approach has the advantage of ensuring the preservation of well-tested cosmological predictions, in particular BBN. 
It is indeed possible to make the axion {\it lighter} than expected. A particle physics model that realises this scenario has been proposed in Ref.~\cite{Hook:2018jle}. 
The model relies on a $Z_N$ symmetry under which $a \to a + \frac{2 \pi f_a}{N}$, and furthermore the axion interacts with $N$ copies of QCD whose fermions transform under $Z_N$ as $\psi_k \to \psi_{k+1}$. Surprisingly, adding up the contributions of all the sectors one finds that cancellations occur in the axion potential with a high degree of accuracy, and as a result, for even $N$ the axion mass gets exponentially suppressed:
\begin{align}
    \label{eq:ZN}
    m_a \to \frac {4 \, m_a}{2^{N/2}} \,, 
\end{align}
while, if $N$ is odd, the axion potential retains the minimum in $\bar\theta =0$. The mass range accessible to FLASH corresponds to $9 \lesssim N \lesssim 13$. This scenario has been further explored in Refs.~\cite{DiLuzio:2021pxd,DiLuzio:2021gos}. A different possibility to generate DM axions with a mass below the canonical window can be engineered within  the framework of the \textit{mirror world}~\cite{Berezhiani:2000gw, Gianfagna:2004je} extended to include the axion, see Ref.~\cite{Giannotti:2005eb}.

\section{Additional models that will be probed with FLASH}
\label{sec:othermodels}

\subsection{Axion-like particles}
\label{sec:ALP}

Along with the QCD axion, other axion-like particles or ALPs can arise in theories of grand unification or quantum gravity~\cite{Svrcek:2006yi, Arvanitaki:2009hb, Arvanitaki:2009fg} and share a phenomenology similar to what discussed so far. Here, an ALP is defined by a negligible axion-gluon coupling in Eq.~\eqref{eq:lagrangian} so that the instantons appearing during the QCD phase transitions do not affect the axion potential and the mass of the particle is independent of the temperature. 
The ALP potential $V(a)$ is usually modeled as a quadratic form to incorporate the explicit breaking origin of the mass term, so that the Lagrangian describing this particle is
\begin{equation}
    \label{eq:potentialALP}
    \mathcal{L} \supset -\frac{1}{4}F_{\mu \nu}\,F^{\mu \nu} + \frac{1}{2}(\partial^\mu a)(\partial_\mu a) - \frac{1}{2}m_a^2a^2 - \frac{1}{4}\,g_{a\gamma\gamma}\,a \tilde F_{\mu \nu}\,F^{\mu \nu}\,.
\end{equation}
Here, an explicit dependence on temperature is not included due to the absence of a strong coupling with the QCD sector. The ALP-photon coupling is also generally expressed in terms of the coupling $g_{a\gamma\gamma}$, similarly to what 
given in Eq.~\eqref{eq:bare_coupling} and for some particle content that justifies the appearance of the $E/N$ ratio.

The same physical scenario described in Sec.~\ref{sec:production} for the cosmological production of  QCD axions holds equally well for an ALP, with the important difference that the potential does not depend on temperature, and the axion abundance in Eq.~\eqref{eq:abundance} when compared to the present DM abundance yields (see e.g.\ Ref.~\cite{Arias:2012az, Visinelli:2017imh})
\begin{equation}
    m_a \approx 5\,\mu{\rm eV} \left(\frac{10^{13}\,{\rm GeV}}{f_a}\right)^4\,\theta_i^{-4}\,.
\end{equation}
When $\theta_i = \mathcal{O}(1)$, the ALP is generally lighter than the QCD axion for a given decay constant. In fact, for $\langle\theta_i\rangle^2 = \pi^2/3$, the ALP mass is in the range of FLASH for $f_a \sim 10^{13}\,$GeV.

\subsection{Scalar dark matter}
\label{ch:sec_dilaton}

So far we have investigated the case in which DM is composed of pseudoscalar particles. The FLASH setup can be used to explore other types of DM candidates such as scalar particles~\cite{Baldeschi:1983mq}, which are motivated in theories of dilaton models~\cite{Cho:2007cy, Choi:2011fy}. Scalar DM can be produced in the early Universe through similar mechanisms as axionic DM, the most prominent for the range of masses of interest for FLASH being VRM.

We consider a system comprising the dilaton field $\phi$ of mass $m_\phi$ and interacting with the EM field strength with a coupling $g_{\phi\gamma\gamma}$, as described by the Lagrangian
\begin{equation}
    \label{eq:dilaton}
    \mathcal{L} \supset -\frac{1}{4}F_{\mu \nu}\,F^{\mu \nu} + \frac{1}{2}(\partial^\mu \phi)(\partial_\mu \phi) - \frac{1}{2}m_\phi^2\phi^2 - \frac{1}{4}\,g_{\phi\gamma\gamma}\,\phi F_{\mu \nu}\,F^{\mu \nu}\,,
\end{equation}
where the last term is the photon-dilaton interaction that can be interpreted as the scalar counterpart of the axion-photon interaction in Eq.~\eqref{eq:lagrangian}. The conversion of $\phi$ in the cavity leads to a signal whose power is analogous to the axion case in Eq.~\eqref{eq:power} upon the replacement $g_{a\gamma\gamma}\rightarrow g_{\phi\gamma\gamma}$ and $m_a\rightarrow m_{\phi}$~\cite{Flambaum:2022zuq}. Due to the different structure between the axion-photon and the dilaton-photon couplings, the coupling to the cavity is given by~\cite{Flambaum:2022zuq}
\begin{equation}
    \label{eq:coupling}
    C_\alpha = \frac{1}{B_0^2V}\,\frac{|\int_V {\rm d}^3{\bf x}\,e^{i{\bf k}\cdot{\bf x}}\,{\bf B}_0\cdot {\bf B}_\alpha|^2}{\int_V {\rm d}^3x\, {\bf B}_\alpha\cdot {\bf B}_\alpha}\,,
\end{equation}
where ${\bf B}_\alpha$ is the magnetic field associated with the resonant mode $\alpha$. Since all of the TM modes as well as the TE$_{010}$ mode possess a vanishing magnetic field along the $z$-direction, we consider here the modes TE$_{011}$ and TE$_{111}$, see Table~\ref{tab:temodes}. The mode TE$_{011}$ is described by the component of the magnetic field along the $z$ direction
\begin{equation}
    B_z = B_0\,J_0(v_0 r/R)\,\sin(\pi z/L)\,,
\end{equation}
where $v_0 \approx 3.832$, while the mode TE$_{111}$ is described by
\begin{equation}
    B_z = B_0\,J_1(v_1 r/R)\,\sin(\pi z/L)\cos\theta\,,
\end{equation}
with $v_1 \approx 1.8412$. In \ref{sec:coupling}, we compute the coupling in Eq.~\eqref{eq:coupling} for the two modes TE$_{011}$ and TE$_{111}$ interacting with the magnetic field of the cavity assuming an isotropic DM distribution for the bosonic wave number $k = mv$ with speed $v \approx 200$\,km/s. Because of the different angular distribution of $B_z$, the two couplings scale differently with the bosonic DM momentum, leading to different estimates as already noted previously~\cite{Flambaum:2022zuq}. In fact, we obtain $C_{011} \approx 5\times 10^{-13}$ and $C_{111} \approx 3\times 10^{-7}$, so that only the bound obtained by considering the TE$_{111}$ mode is consistently competitive with the existing bounds.

\begin{figure}[!ht]
    \begin{center}
    \includegraphics[width=\linewidth]{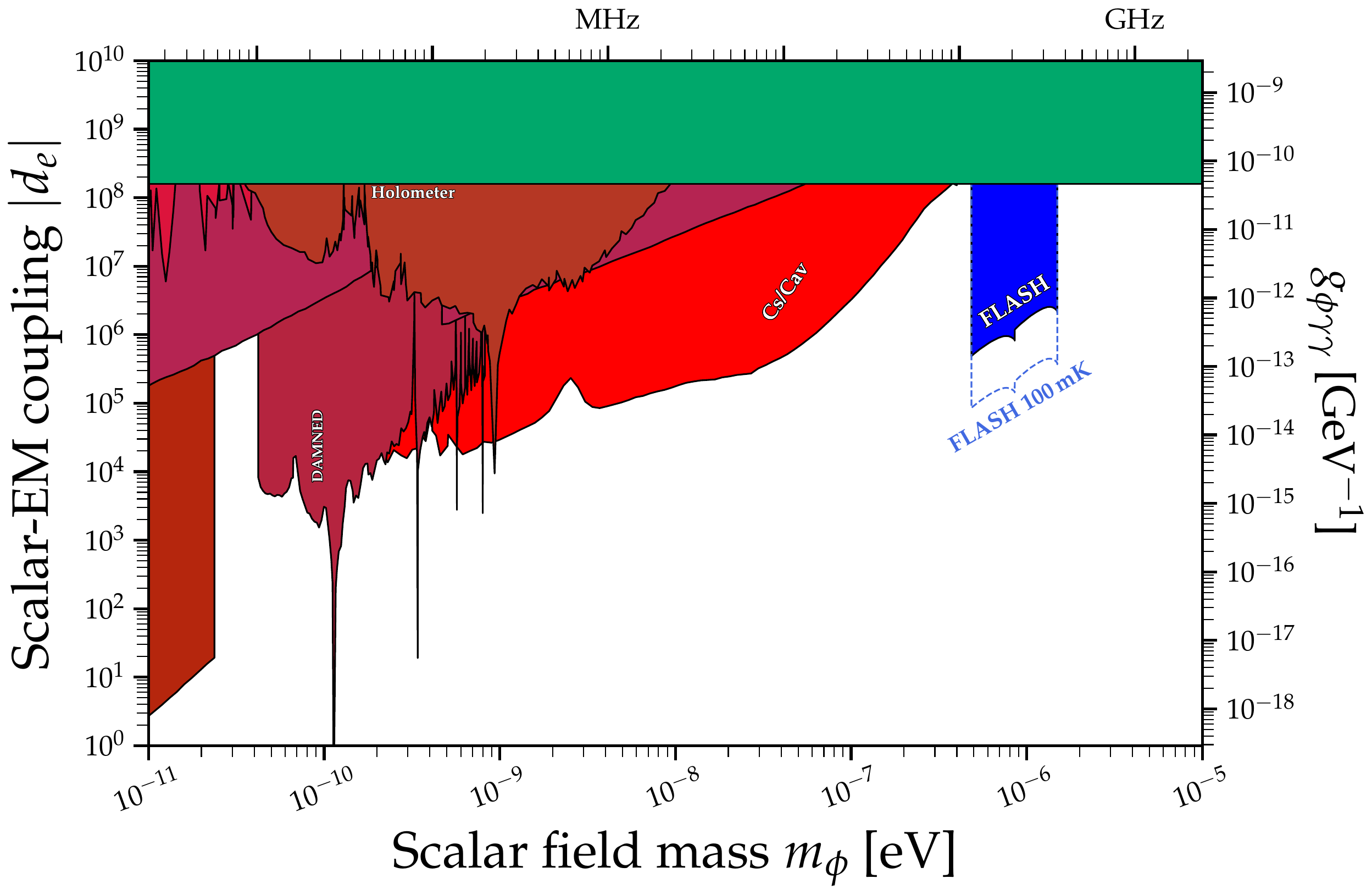}
    \caption{The predicted sensitivity in FLASH for the coupling between a scalar particle and the photon $|d_e|$ as expressed in Eq.~\eqref{eq:dilaton}, for the setup adopted for the first stage (filled blue area) and once the cryogenic has been improved to reach $T_{\rm sys} = 100$\,mK (dashed blue line). Also shown are current laboratory bounds, see Ref.~\cite{AxionLimits} for details.}
    \label{fig:ScalarPhoton}
    \end{center}
\end{figure}

Figure~\ref{fig:ScalarPhoton} shows the forecast on the constraint for the scalar field coupling with the photon, once the coupling obtained for the axion field has been rescaled by a factor $\sqrt{C_{111}/C_{010}} \approx 10^{-3}$. This is merely an estimate based on the simulations for the axion reach, since the TE$_{111}$ mode has a different quality factor and frequency range (see Tab.~\ref{tab:temodes}). A proper analysis that considers the coupling of the mode to the cavity is required and will be carried out along with the data analysis.

The bound has been expressed in terms of the dimensionless quantity $|d_e| \equiv \sqrt{2}M_{\rm Pl}g_{\phi\gamma\gamma}$ that is relevant for dilaton models~\cite{Damour:1994zq, Damour:2010rm}. Also shown are the current bounds placed by laboratory searches for $|d_e|$ and reported in Ref.~\cite{AxionLimits}. Future experiments involving a mechanical resonator made of a single crystal can lead to competitive bounds in the mass range of interest for FLASH~\cite{Manley:2019vxy} Note, that fifth-force and equivalence principle searches are generally stronger than laboratory constraints~\cite{Wagner:2012ui, Hees:2016gop, Berge:2017ovy}. For an update of the the bounds and forecasts on the coupling of scalar particles in various mass ranges see the recent review in Ref.~\cite{Antel:2023hkf}.

\subsection{Chameleons}
\label{ch:sec_chameleons}

The ``chameleon'' was first introduced to potentially explain the present accelerating expansion of the Universe through a variable effective mass that depends on the ambient energy density.\footnote{However, see Ref.~\cite{Wang:2012kj} for a no-go theorem related to self-acceleration and for possible caveats.} The chameleon field $\phi$ is generally characterized by an effective interaction whose strength is a function of the local environment~\cite{Khoury:2003aq}, with a ``screening'' potential modeled as~\cite{Brax:2004qh, Vagnozzi:2021quy}
\begin{equation}
    \label{eq:effective_potential}
    V_{\rm eff}(\phi) = V(\phi) + \exp\left(\frac{\beta_m\phi}{M_{\rm Pl}}\right)\rho_m + \frac{1}{4}\exp\left(\frac{\beta_\gamma\phi}{M_{\rm Pl}}\right)F^{\mu\nu}F_{\mu\nu}\,.
\end{equation}
Here, $\beta_m$ and $\beta_\gamma$ are couplings to the density in matter $\rho_m$ and photons, respectively, and $V(\phi) \propto \phi^{-n}$ is the self-interacting potential of the chameleon in the absence of other couplings, which depends on the index $n \geq 0$.

The effective mass $m_{\rm eff}$ associated with the potential in Eq.~\eqref{eq:effective_potential} depends on the couplings and on the relative energy content in matter and radiation. It is therefore possible that the chameleon mass in the walls of the cavity is higher than the mass inside the cavity itself, effectively trapping the field along with the magnetic field~\cite{Jaeckel:2006xm, Brax:2007hi, Ahlers:2007st}. In the interspace within the cavity walls, the chameleon field follows the Klein-Gordon equation~\cite{Ahlers:2007st}
\begin{equation}
    \left(\frac{{\rm d}^2}{{\rm d}t^2} - \nabla^2 + m_{\rm eff}^2\right)\phi = \frac{\beta_m}{M_{\rm Pl}}\,{\bf B}\cdot ({\bf \nabla} \times {\bf A})\,,
\end{equation}
where ${\bf A}$ is the potential of the propagating photon. Once the chameleon field is trapped inside the resonant cavity, it experiences a delayed decay through the so-called afterglow effect~\cite{Ahlers:2007st, Gies:2007su}. To reconstruct the signal produced inside the cavity, we derive the field equations from Eq.~\eqref{eq:effective_potential} as~\cite{Ahlers:2007st}
\begin{equation}
    \Box A = \frac{\beta_m}{M_{\rm Pl}}\,{\bf \nabla}\phi\times {\bf B}\,,
\end{equation}
with the condition that we observe a standing wave instead of a moving wave as in the DM case.

\begin{figure}[!ht]
  \begin{center}
    \includegraphics[width=.7\linewidth]{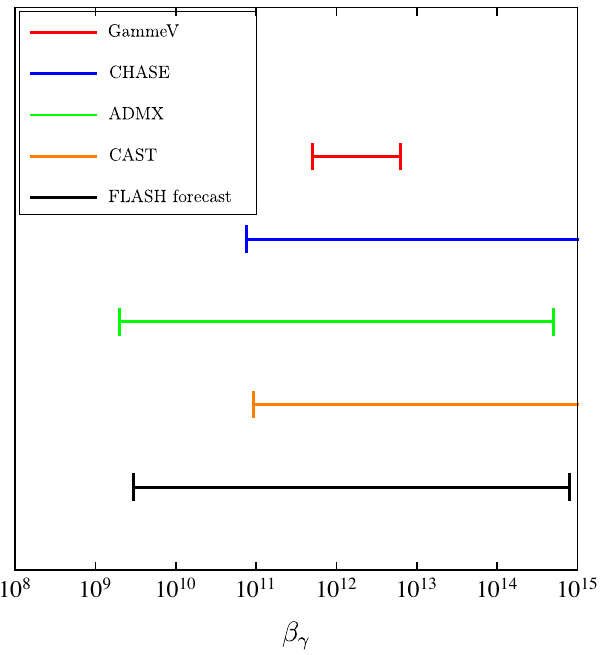}
    \caption{Forecast reach for the search of the chameleon-photon coupling with FLASH, for the setup described in the text (black line). Also shown are the results obtained by GammeV~\cite{GammeV:2008cqp} (red line), CHASE~\cite{Steffen:2010ze, Upadhye:2012ar} (blue line), ADMX~\cite{ADMX:2010ndb} (green line), and CAST~\cite{CAST:2015npk} (orange line).}
    \label{fig:chameleons}
  \end{center}
\end{figure}

We estimate the potential reach of FLASH for the coupling of the chameleon with the photon $\beta_\gamma$, following the procedure outlined in Ref.~\cite{ADMX:2010ndb} in which the TE$_{011}$ cavity mode is excited by an external source of power $P_{\rm in} = 0.5\,$mW for a period $t_0 = 10\,$min, with a sweeping time $\tau = 10\,$min in Eq.~\eqref{eq:snr} and a noise temperature $T_{\rm sys} = 4.5\,$K. The natural frequency of the TE$_{011}$ mode resonating in the large cavity setup is $f_{\rm res} \approx 214.5\,$MHz, with a quality factor $Q \approx 1.3\times 10^6$ (see Tab.~\ref{tab:temodes}). Following the computation related to the expression in Eq.~\eqref{eq:coupling}, we fix the coupling of the cavity $C_{011} \approx 0.005$ as shown in \ref{sec:coupling}. This leads to the forecasts at one sigma level:
\begin{equation}
    3\times 10^9 \lesssim \beta_\gamma \lesssim 8\times 10^{14}\,,
\end{equation}
which overlaps with previous searches that already exclude the coupling for cosmic chameleons in some ranges, including GammeV~\cite{GammeV:2008cqp}, GammeV-CHASE (here CHASE)~\cite{Steffen:2010ze, Upadhye:2012ar}, ADMX~\cite{ADMX:2010ndb}, and CAST~\cite{CAST:2015npk}. These results have been summarized in Fig.~\ref{fig:chameleons}.

\subsection{Hidden photon dark matter}
\label{ch:sec_dark-pho}

This subsection is devoted to the simplest case of \emph{hidden photon} (HP)~\cite{Okun:1982xi, Holdom:1985ag, Foot:1991kb} dark matter.\footnote{These particles have also been dubbed ``dark photons'' or ``paraphotons''.} Although several extensions are possible and have been proposed in the literature the advantage of the realization discussed here is that the experimentally constrained signal is determined by two parameters~\cite{Arias:2012az}: the kinetic mixing term $\chi$ and the mass of the HP itself $m_{\gamma{\,\prime}}$. The dimensionless mixing parameter $\chi$ can be probed experimentally as a function of $m_{\gamma{\,\prime}}$, and is generally expected to be $\chi\ll 1$. 
For this reason, we work in this limit hereafter.

Before delving into the discussion it should be mentioned that $\mathcal{O}(1)$ fraction of DM can be obtained from inflationary fluctuations in the form of hidden photon for~\cite{Graham:2015rva}
\begin{equation}
    \label{eq:mdarkphotonDM}
    m_{\gamma{\,\prime}} \simeq 10^{-5}{\rm\, eV} \,\left( 10^{14}{\rm\,GeV}/H_I\right)^4\,,
\end{equation}
depending on the scale of inflation $H_{I}$. As mentioned before, FLASH will be able to constrain the mass scales of hypothetical particles in the range around $\sim 10^{-6}{\rm\,eV}$, therefore close to values expected in Eq.~\eqref{eq:mdarkphotonDM} for a narrow range of $H_I$. To our knowledge there is no consensus on a formation model leading to the production of the right abundance of dark matter. Therefore, in the following we will simply assume this to be the case and we remain agnostic on the exact mechanism of production (see Ref.~\cite{Caputo:2021eaa} for a review).

\begin{figure}[!ht]
    \begin{center}
    \includegraphics[width=0.8\linewidth]{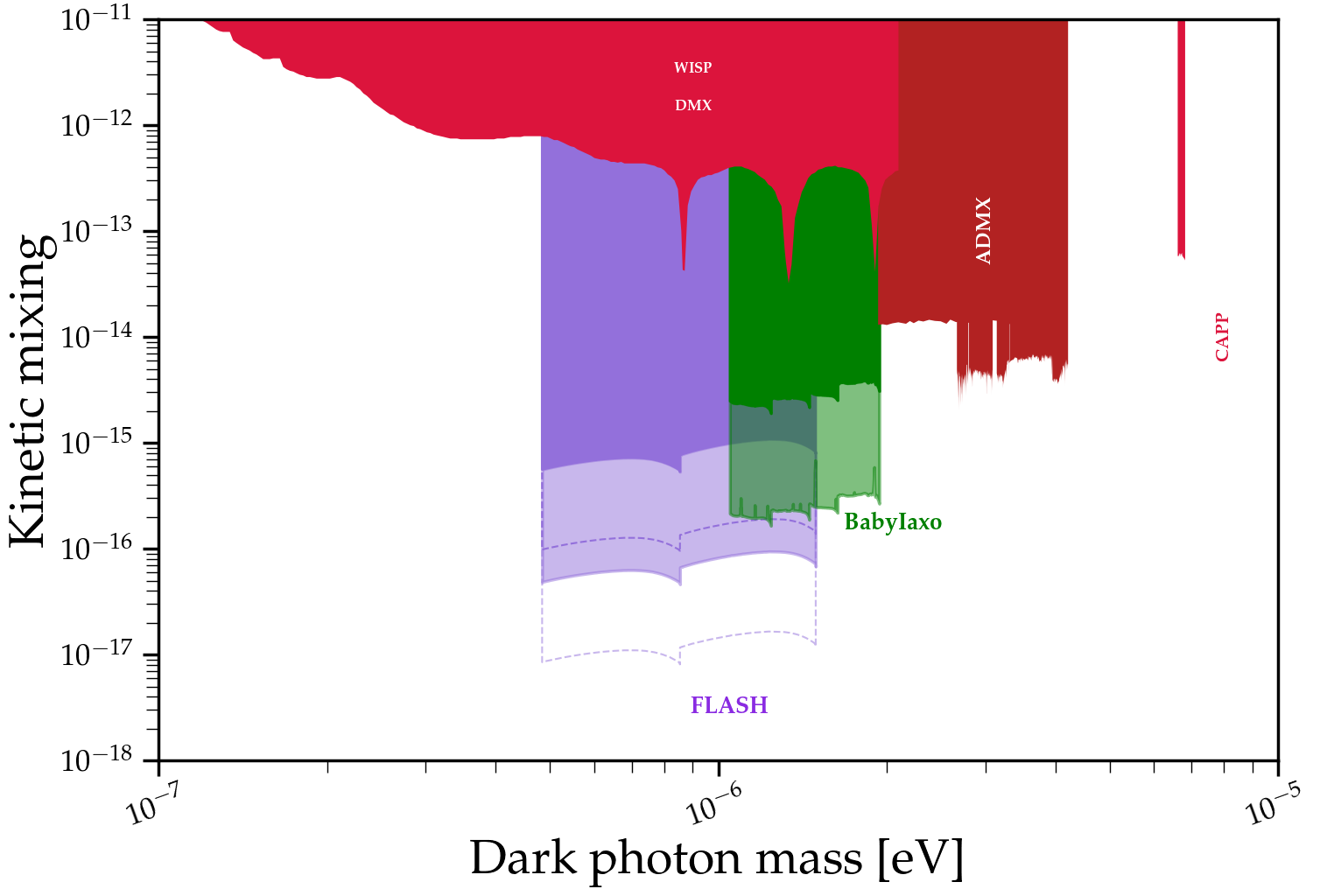}
    \caption{Compilation of current constraints and projected sensitivities to photon-HP mixing of current and planned experiments in the $m_{\gamma^{\,\prime}}- \chi$ plane. The predicted sensitivity for FLASH is reported with full lines of different purple shades. The shaded dark purple region, represents the predicted FLASH sensitivity in the scenario of highly directional HP field ($\cos^2 \theta = 0.0025$). The shaded light purple region reports the predicted FLASH sensitivity in the scenario of the isotropic distribution of the HP field ($\cos^2 \theta = 1/3$). The projections from ~\cite{Ahyoune:2023gfw} labelled `babyIAXO' in green follow the same color-scheme.
    The dashed lines represent the FLASH forecast sensitivities when operating the haloscope at a temperature of 100\,mK. Dashed dark purple line: highly directional HP field; Dashed light purple line: isotropic distribution of the HP field. Image realized with the code in Ref.~\cite{AxionLimits}.}
    \label{fig:chivsm}
    \end{center}
\end{figure}

The HP field $X_\mu$ describes a hidden $U(1)$ symmetry group that mixes with the photon through a Lagrange density of the form~\cite{Redondo:2010dp}
\begin{equation}
    \mathcal{L} \supset - \frac{1}{4} F_{\mu\nu}\,F^{\mu\nu} - \frac{1}{4} X_{\mu\nu}\,X^{\mu\nu} + \frac{\chi}{2} X_{\mu\nu}\,F^{\mu\nu} + \frac{1}{2} m^2_{\gamma^{\,\prime}}X_\nu\,X^\nu- j_{\mathrm{em}}^\mu\,A_\mu\,,
    \label{hp:eqL1}
\end{equation}
where $X_\mu$ is the field describing the HP with field strength tensor $X_{\mu\nu}$ and $j_{\mathrm{em}}^\mu$ is the electromagnetic current. The Lagrangian in Eq.~\eqref{hp:eqL1} leads to a possible decay channel of the HP into three photons~\cite{Pospelov:2008jk}. Demanding that the HP is a stable relic on cosmological timescales leads to the bound \begin{equation}
    m_{\gamma{\,\prime}} \,(\chi^2\alpha_{\rm EM})^{1/9}\lesssim 1{\rm\,keV}\,,    
\end{equation}
which is clearly satisfied for the range of masses of interest in FLASH, see Fig.~\ref{fig:chivsm}. This motivates the search for the HP as the cosmic DM.

Note, that the microwave resonant cavity experiments searching for axion can also be used to probe the photon-HP mixing when operated without the magnetic field~\cite{Arias:2012az}, which is unnecessary in this case. The equation of motion for the photon field $A^\mu$ 
\begin{equation}
    \partial_{\mu}\partial^{\mu} A^{\nu} = \chi\,m^2_{\gamma{\,\prime}}\,X^{\nu}\,,
    \label{hp:eqM}
\end{equation}
implies that HP could source ordinary photon. In particular, the power emission of the cavity is related to the energy stored and the quality factor of the cavity, and it is given by~\cite{Arias:2012az, Ghosh:2021ard}
\begin{equation}
    P_{\mathrm{sig}} = \left(\chi^2 \,\rho_{\gamma{\,\prime}}\,\cos^2 \theta\right)\times \left(\omega_c\, Q_L\,\frac{\beta}{1+\beta}\, V\, C_{nml}\right)\,,
    \label{hp:eq4}
\end{equation}
where $Q_L$ is the quality factor, $V$ the volume of the cavity and $\theta$ the direction between the hidden photon field and the magnetic field.

Assuming the energy density of the HP to be equal to the DM density, the same quality factor used for the axion search, and the volume of the cavity equal to $V_{\rm Large}$ for the large phase of the experiment or $V_{\rm Small}$ for the small phase, Eq.~\eqref{hp:eq4} gives the expected power in the cavity that enters the SNR in Eq.~\eqref{eq:snr}. This leads to a constraint of the HP as the DM with the FLASH microwave cavity. The large dimensions of the cavity significantly enhance the sensitivity for HP. The predicted sensitivity for FLASH is computed inserting in Eq.~\eqref{eq:snr} the $P_{\mathrm{sig}}$ obtained from Eq.~\eqref{hp:eq4}. Two scenarios are possible, depending on the relative orientation of the cavity with respect to an a priori unknown direction of the HP field. This is represented by the factor $\cos^2 \theta$ in Eq.~\eqref{hp:eq4}. Assuming that all directions in space are equally likely and demanding that the real value is bigger with 95\% probability leads to the value $\cos^2 \theta = 0.0025$~\cite{Ghosh:2021ard, Caputo:2021eaa},\footnote{Here we use the value $0.0025$ for comparison with previous experiments. A truly 90\% C.L.\ can only constraint the product $\chi\cos\theta$.} while an average over all possible directions gives $\cos^2 \theta = 1/3$.

Fig.~\ref{fig:chivsm} shows the predicted HP detection sensitivity of the FLASH experiment for the setup discussed above and for $\cos^2 \theta = 0.0025$ (shaded dark purple area) or $\cos^2 \theta = 1/3$ (shaded light purple area), the latter being a more optimistic forecast. Also shown are the sensitivities expected in an upgrade setup of FLASH with an improved cryogenics able to reach the noise temperature $T_{\rm sys} = 100\,$mK (dashed lines). Finally, we also show a compilation of current constraints on and sensitivities of planned experiments to photon-HP mixing in the $m_{\gamma^{\,\prime}} - \chi$ plane~\cite{AxionLimits}. 
Even in the most conservative scenario and with the planned setup, FLASH will be able to probe the HP coupling orders of magnitude below.

\subsection{Detection of gravitational waves}
\label{ch:sec_GW}

The detection of high frequency gravitational waves (HFGW) in the MeV-GeV range has recently attracted interest with many proposals currently under development, moved by the theoretical considerations that predict various exotic sources to emit at such high frequencies~\cite{Aggarwal:2020olq}. Potential sources of HFGWs are generally divided into two categories, namely sources of cosmological origin produced before recombination and sources of astrophysical origin. A cosmological background of stochastic HFGWs is expected from various sources, including the primordial thermal plasma~\cite{Ringwald:2020ist} or a first order phase transition in the dark sector~\cite{Addazi:2017nmg}. When of cosmological origin, the HFGW strain $h_0$ as a function of the HFGW frequency $\nu_c$ is constrained by BBN considerations and CMB data to be~\cite{Cyburt:2015mya,Domcke:2022rgu, Domcke:2023bat}
\begin{equation}
    h_0\lesssim 10^{-29}\,\left(100 \,{\rm MHz}/\nu_c\right)\,\Delta N_{\rm eff}^{1/2}\,,
\end{equation}
several orders of magnitudes below the expected reach of cavity searches~\cite{Berlin:2021txa}. For this, in the following we consider astrophysical sources in the late Universe at such high frequencies, whose origin likely requires new physics.

\subsubsection{Gravitational waves from compact objects}

Besides compact objects of astrophysical origin such as black holes (BH) and neutron stars, exotic configurations in the form of primordial BHs or boson stars formed in the earliest stages of the Universe could also be present today. As an example, primordial BHs (PBH) are too light to be explained by known stellar dynamics and hence, if detected, their origin would require new physics. 
A possibility is that they are formed in the early Universe, hence the name primordial (see Ref.~\cite{Escriva:2022duf} for a recent review). 
The exact details of the formation scenario is currently unknown. Historically, they have been conjectured to follow from inflationary overdensities~\cite{Hawking:1971ei,Zeldovich:1967lct,Carr:1974nx}. 
However, these scenarios require significant tuning in order to produce a sizeable amount of dark matter, see e.g.\ Ref.~\cite{Cole:2023wyx} for a recent discussion. Over the past years, mechanisms that avoid such issues have been proposed. For example, PBHs could form due to the preheating dynamics~\cite{Martin:2019nuw}, or from the confinement of heavy quarks~\cite{Dvali:2021byy}. 
Regardless of the specific formation scenario, PBHs could constitute a substantial fraction of dark matter for masses of the order of $\left(10^{-15}-10^{-11}\right)\,M_{\odot}$. 
Heavier PBHs as the main DM constituents are excluded by severe lensing constrains, although a small contribution is still allowed~\cite{Carr:2020gox}.

The GWs emitted by binary configurations of compact objects could possess the frequency range and the strain required to be detectable with present or near-future technologies. Consider two compact objects of similar mass $M$ and size $R$, each of compactness $\mathrm{C} \equiv GM/R$, forming a system of total mass $M_{\rm tot} \approx 2M$. The frequency of the emitted GW spectrum at the end of the inspiral phase when the innermost stable circular orbit (ISCO) is occupied is $\nu_{\rm ISCO}$. For the case of two coalescing PBHs, the GW frequency at the ISCO is~\cite{Giudice:2016zpa}
\begin{equation}
	\label{eq:fISCO}
	\nu_{\rm ISCO} = \frac{\mathrm{C}^{3/2}}{3\sqrt{3}\pi G M_{\rm tot}}\,,
\end{equation}
so that a signal in the bandwidth $\mathcal{O}(100\,{\rm MHz})$ and for the compactness of a BH, $\mathrm{C}=0.5$, is expected for PBHs in the mass range $M_{\rm PBH} \sim 10^{-5}\,M_{\odot}$. The GW strain sourced by a merging event of total PBH mass $M_{\rm tot}$ is~\cite{Maggiore:2007ulw}
\begin{equation}
    \label{eq:PBHstrain}
    h_0 \simeq 9.77\times 10^{-28}\left(\frac{\rm kpc}{d}\right)\left(\frac{M_{\rm tot}}{10^{-8}M_\odot}\right)^{\frac{5}{3}}\left(\frac{\nu}{100\,\rm MHz}\right)^{\frac{2}{3}}\,,
\end{equation}
where the distance $d$ between the Earth and the source is found by demanding that at least one merger event per year occurs within $d$~\cite{Franciolini:2022htd}. The rate of merger events is obtained following the analyses in Refs.~\cite{Raidal:2018bbj, Domcke:2022rgu}.

\subsubsection{FLASH forecast reach for gravitational wave detection}

A resonant cavity designed to probe the conversion of axions into photons can also be used to detect gravitational waves through a graviton-photon conversion via the inverse Gertsenshtein effect~\cite{Gertsenshtein:1962} (see also Refs.~\cite{Zeldovich:1973,Palessandro:2023tee}). 
In its simplest formulation, the coupling of the photon with gravity is described by the Maxwell-Einstein action,
\begin{equation}
    S = \int {\rm d}^4x \sqrt{-g}\left( - \frac{1}{4}g_{\mu\alpha}g_{\nu\beta}F^{\mu\nu}F^{\alpha\beta}\right)\,,
\end{equation}
where $g_{\mu\nu}$ is the space-time metric with determinant $g$. The expansion of the metric to first order around a flat background as $g^{\mu\nu} = \eta^{\mu\nu} + h^{\mu\nu}$, with $|h^{\mu\nu}| \ll 1$, leads to a coupling between a GW of strain $h_0$ and frequency $\omega_g$ in an external magnetic field $B_0$ as $\propto h_0 E B_0$, where $E$ is the electric field. Including the effects of the source requires adding a Breit-Wigner distribution of the form~\cite{Hong:2014vua, OHare:2017yze}
\begin{equation}
    F(\omega) = \frac{1}{(\omega-\omega_c)^2 + \frac{1}{4\tau^2}}\,,
\end{equation}
where the relaxation time is defined as the time spent by the binary system within the frequency range $\Delta \nu$,\footnote{Note, that this definition differs from other work in the literature~\cite{Franciolini:2022htd}, where $N_{\rm cycle}$ is defined in terms of the frequency range $\Delta \nu \sim \nu_c$.}
\begin{equation}
    \tau = \frac{\Delta \nu}{\dot\nu_c} = \frac{2\pi N_{\rm cycle}}{\omega_c}\,.
\end{equation}
A quality factor describing the coherence of the source can then be defined as $Q_S = 2\pi N_{\rm cycle}$, analogously to the quantity $Q_a$ introduced earlier for the axion. An electric signal of the same frequency is then produced in the cavity with the signal power~\cite{Berlin:2021txa}
\begin{equation}
    P_{\rm sig} = \frac{1}{2}Q_{\rm eff}\,\omega_c^3\, V^{5/3}\,\left(\eta h_0B_0\right)^2\,,
\end{equation}
where $Q_{\rm eff} = \min(Q_L, Q_S)$.

In the limit of an infinitely coherent source, the sensitivity of the cavity search to GW detection estimated through Eq.~\eqref{eq:snr} leads to~\cite{Berlin:2022hfx}
\begin{eqnarray}
    \label{eq:strainberlin}
    h_0 &\gtrsim&  9\times10^{-22}\left(\frac{200{\rm \,MHz}}{\omega_g/2\pi}\right)^{\frac{3}{2}}\left(\frac{0.1}{\eta}\frac{\rm{T}}{B_0}\right)
    \left(\frac{4{\rm\, m}^3}{V}\right)^{\frac{5}{6}}\left(\frac{2\times10^5}{Q_L}\frac{T_{\rm sys}}{5 \rm \,K}\right)^{\frac{1}{2}}
    \\\nonumber
    &&\left(\frac{\Delta\nu}{1{\rm \,kHz}}\frac{2{\rm \,min}}{t_{\rm int}}\right)^{\frac{1}{4}}\,,
\end{eqnarray}
where the expression accounts for an experimental setup similar to that of FLASH, corresponding to the FINUDA magnet with a nominal magnetic field $B_0 = 1.1\,$T and a cavity volume $V_{\rm Large} = 4.15{\rm\,m^3}$. We have also set the measurement integration time $t_{\rm int} = 2\,$min with the signal bandwidth (BW) equal to the cavity bandwidth $\Delta \nu = 1\,$kHz and the cavity-GW coupling coefficient $\eta = 0.14$ for the mode TM$_{012}$. The quality factor for the mode is expected to be $Q_0 \simeq 7\times 10^5$. For comparison, the sensitivity of ADMX could be in the range $h_0 \sim 10^{-22}$, where the difference comes from the fact that although the FINUDA magnet has a smaller magnetic field, it has a much larger effective volume than the ADMX magnet. In a realistic setup where the coherence of the source is limited by $Q_S$, Eq.~\eqref{eq:strainberlin} holds when replacing $Q_L \to Q_{\rm eff}$ and $t_{\rm int} \to \min(t_{\rm int}, \tau)$; moreover, we demand that the source satisfies $N_{\rm cycle} \gtrsim 1$. For $N_{\rm cyc}\gtrsim Q_L$ matched filtering with a table of expected waveforms could also be implemented allowing to substitute $\Delta\nu \sim 1/t_{\rm int}$, so that the sensitivity in Eq.~\eqref{eq:strainberlin} would scale differently~\cite{Berlin:2022hfx}:
\begin{eqnarray}
    \label{eq:strainberlinmatched}
    h_0 &\gtrsim&  5\times10^{-23}\left(\frac{200{\rm \,MHz}}{\omega_g/2\pi}\right)^{\frac{3}{2}}\left(\frac{0.1}{\eta}\frac{\rm{T}}{B_0}\right)
    \left(\frac{4{\rm\, m}^3}{V}\right)^{\frac{5}{6}}
    \\\nonumber
    &&\times \left(\frac{2\times10^5}{Q_L}\frac{T_{\rm sys}}{5 \rm \,K}\right)^{\frac{1}{2}}
    \left(\frac{2{\rm \,min}}{t_{\rm int}}\right)^{\frac{1}{2}}\,.
\end{eqnarray}

By comparing the theoretical expectation in Eq.~\eqref{eq:PBHstrain} with the reach in Eq.~\eqref{eq:strainberlin}, it could be expected that an event originating within a distance $d\lesssim 1$\,kpc would fall within the reach of FLASH. However, the in-spiraling binary that sources the GW signal is not as coherent as DM axions are. This results in an effective limitation of the source resonating with the detector.
Similarly, the bandwidth also depends on how fast the system spans a given frequency range. It follows that $\Delta \nu \equiv \nu_c/Q_{\rm eff}$~\cite{Franciolini:2022htd}.

\begin{figure}[!ht]
  \begin{center}
\includegraphics[width=.7\linewidth]{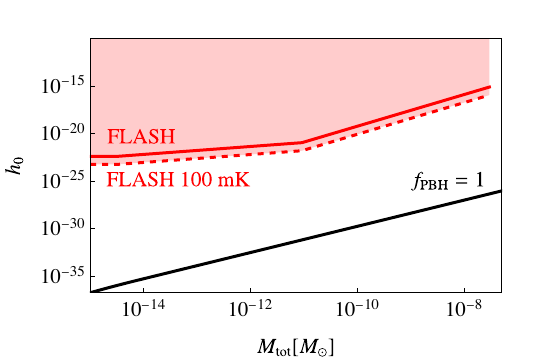}
    \caption{Comparison of FLASH reach for different PBH masses. The red line correspond to the reach of FLASH according to the values described below Eq.~\eqref{eq:strainberlin}. The dashed red line corresponds to the reach with improved cryogenics. 
    The black line correspond to the expected gravitational wave signal in case of a monochromatic spectrum at $f_{\rm PBH}=1$. The frequency of the signal is fixed to $\nu_c = 200$\,MHz.}
    \label{fig:GWFLASHreach}
  \end{center}
\end{figure}

Fig.~\ref{fig:GWFLASHreach} shows the reach of the GW strain in Eq.~\eqref{eq:strainberlin} as a function of the binary mass $M_{\rm tot}$ for the planned FLASH setup (red solid line) and for an upgraded cryogenic system to $T_{\rm sys}= 0.15\,\rm K$ (red dashed line). Also shown is the signal generated by a merger event as in Eq.~\eqref{eq:PBHstrain} for $f_{\rm PBH}=1$, assuming that at least one event per year is observed~\cite{Franciolini:2022htd, Domcke:2022rgu}. For simplicity, a monochromatic spectrum has been assumed at each value of $M_{\rm tot}$. 
The rate of mergers in the range of interest is also enhanced by the galactic overdensity~\cite{Pujolas:2021yaw}. Theoretical uncertainty arises from several unknowns in the early Universe. For example, as shown in~\cite{Franciolini:2022htd}, the presence of non-Gaussianity in the inflationary density perturbation can lead to a significant increase of the rate up to roughly two orders of magnitude.

In Fig.~\ref{fig:GWFLASHreach} the frequency is fixed at $200\,\rm MHz$, which is relevant for the FLASH setup. Mergers with black holes heavier than about $10^{-7}M_{\odot}$ produce a signal resonating within the cavity for less than one cycle therefore requiring a different search strategy. 
For lighter black holes, the lack of signal coherence - in the limit where $N_{\rm cycles}\ll Q_{L}$ - causes a significant decrease of the sensitivity reach, down to masses of order $10^{-11}M_{\odot}$. At a first glance, the reach of FLASH might seem much worse than the estimates proposed in Refs.~\cite{Berlin:2022hfx,Franciolini:2022htd,Domcke:2022rgu}. However, previous results follow from the assumption of a sufficiently coherent source within the frequency bandwidth, which is not the case for merging PBHs as already discussed above.
Finally, for black holes heavier than about $10^{-14}M_{\odot}$ a drop in the sensitivity is also caused by the fact that the duration of the signal in the relevant bandwidth is shorter than the chosen integration time of order minute. Slightly different values, but of comparable order are obtained within the frequency range of FLASH experiment. 


\section{RF cavity design and tuning}
\label{sec:RF}

\begin{figure}[!ht]
\begin{center}
	\includegraphics[width=0.6\linewidth]{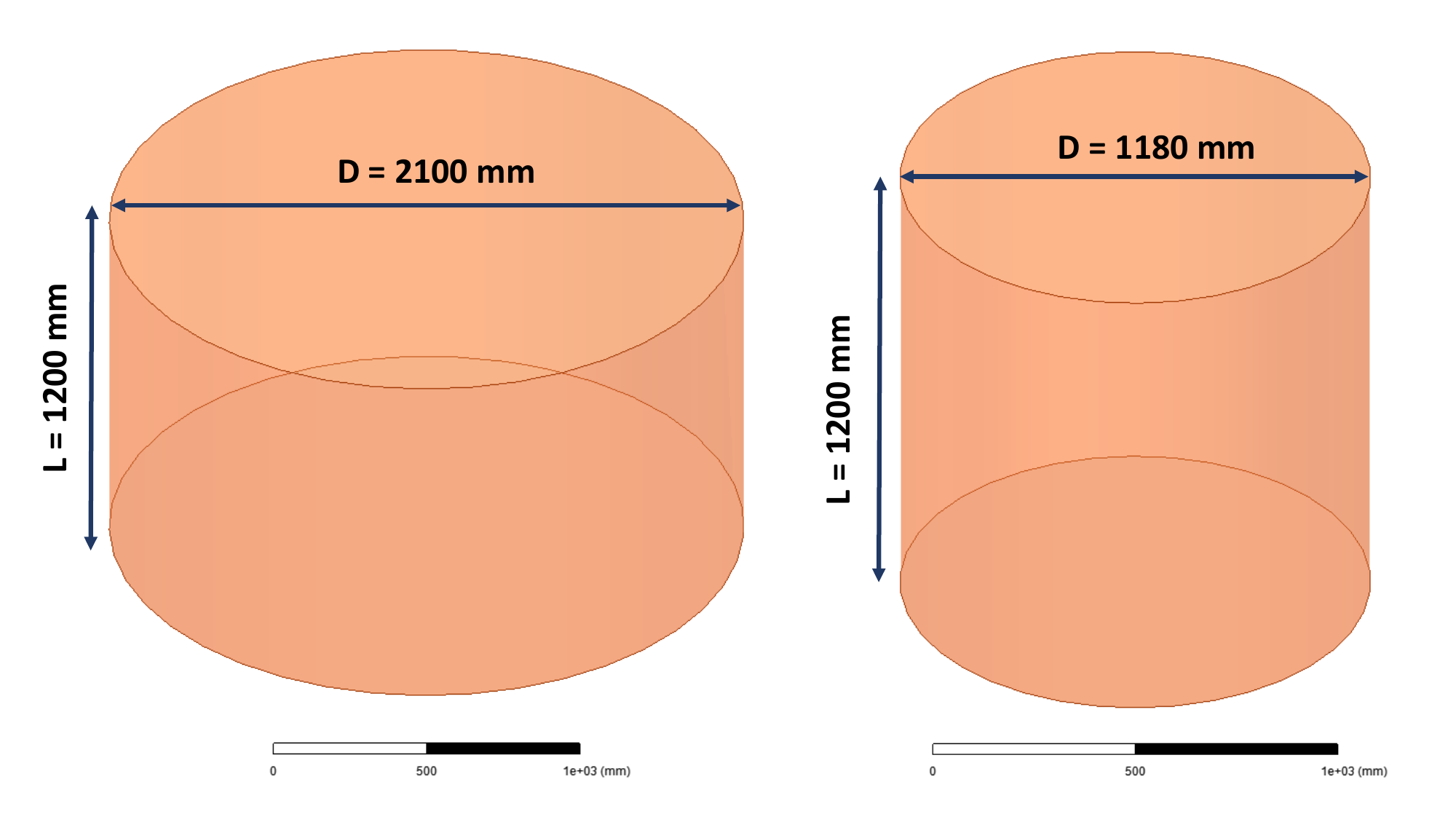}
	\caption{Schematics of the two cavities (without tuning system) proposed to cover the frequency range of resonant frequencies (117 - 360)\,MHz.}
	\label{fig:cav-sket}
\end{center}
\end{figure}

The goal of the FLASH experiment is to cover the range $\nu_c \sim$ (117 - 360)\,MHz by tuning its resonant frequency that operates on the mode TM$_{010}$ for the axion search. Since this wide range of tuning frequencies cannot be covered in a single setup, two cylindrical resonant cavities of different volumes have been considered, each with its tuning system. The larger cavity has a length of 1200\,mm and a radius of 1050\,mm for an inner volume $V_{\rm Large} = 4.15{\rm\,m^3}$ and it is expected to operate in the frequency range $\sim$(117 - 206)\,MHz during the first phase of the operation (FLASH Low Frequency or LF). The second cavity has a length of 1200\,mm and a radius of 590\,mm for an inner volume $V_{\rm Small} = 1.31{\rm\,m^3}$ and will explore the range $\sim$(206 - 360)\,MHz during the second phase (FLASH High Frequency or HF). The two setups overlap over a few MHz. The two cavities have different diameter and are schematically represented in Fig.~\ref{fig:cav-sket}.

The resonant cavity will be made of oxygen-free high thermal conductivity copper (OFHC), a type of copper which may show a residual resistance ratio (RRR) that varies between 50 and 700. 
Assuming RRR = 50, we simulated the quality factor $Q$ and the form factor $C_{010}$ of the TM$_{010}$ resonant mode of the FLASH resonant cavities with the ANSYS-HFSS code.\footnote{\href{https://www.ansys.com/}{https://www.ansys.com/}} Simulations predict values for the quality factor that vary from about $3.8\times 10^5$ to $5.7\times 10^5$, while the form factor varies from about $0.63$ to $0.75$. These results are used to determine the FLASH sensitivity to QCD galactic DM axions and hidden photons.

\begin{figure}[!ht]
\begin{center}
	\includegraphics[width=0.8\linewidth]{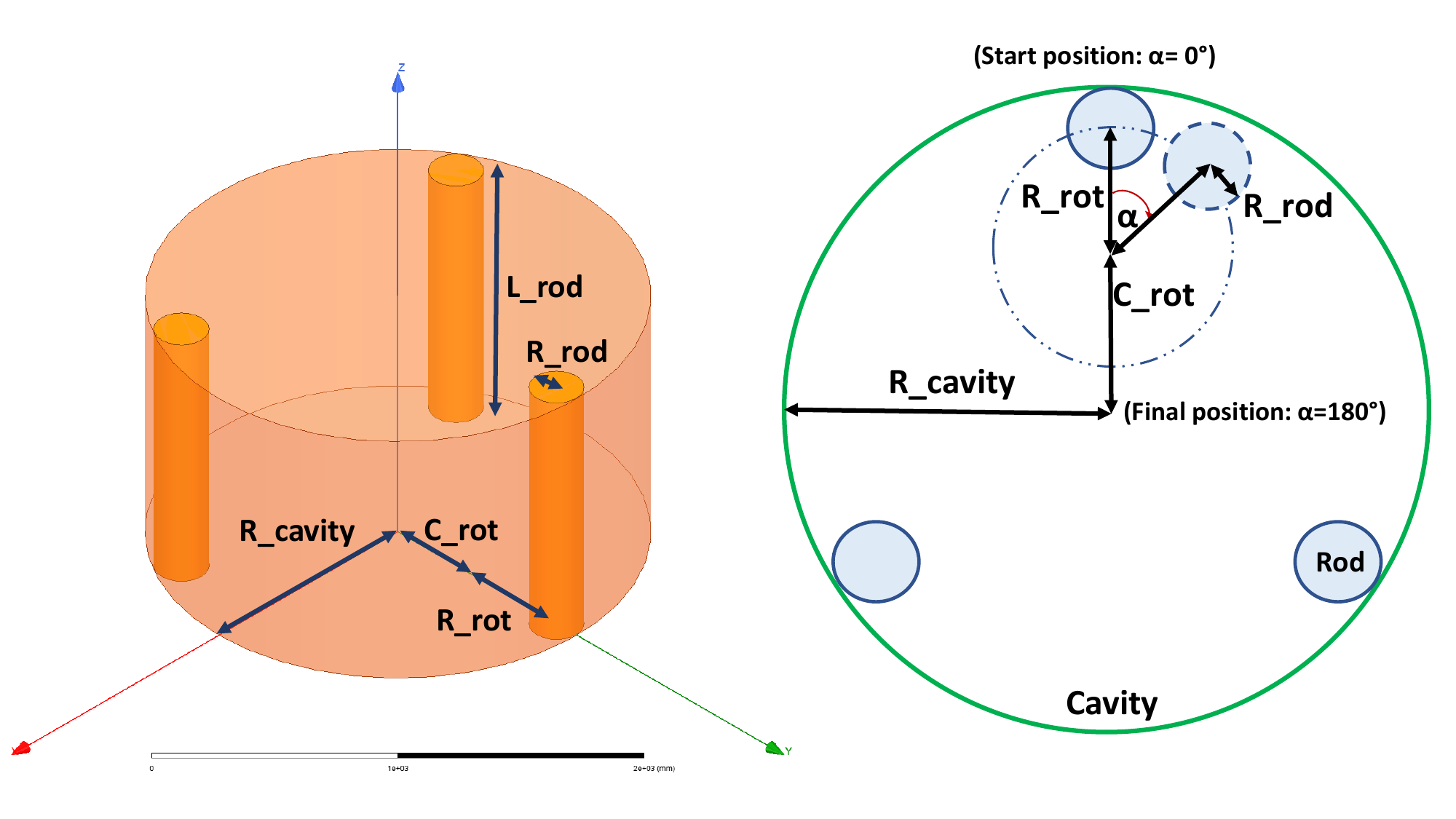}
	\caption{Sketch of the cavity with tuning system.}
	\label{fig:cavity}
\end{center}
\end{figure}

The tuning system is based on the use of metallic movable rods, similarly to what has been adopted by the ADMX collaboration~\cite{ADMX:2001dbg, ADMX:2001nej, Lee:2015aht, Stern:2015kzo}. After a preliminary investigation in which different possible configurations have been explored in terms of number of rods, dimensions and positions, the optimized case of three rods has been selected as represented schematically in Fig.~\ref{fig:cavity}. The final parameters are reported in the Table~\ref{tab:param} for both cavities. Each rod of radius $R_\mathrm{rod}$ rotates by an angle $\alpha$ around a center $C_\mathrm{rot}$, moving towards the center of the cavity on a circular trajectory of radius $R_\mathrm{rot}$. The rotation of the rod allows for the tuning of the mode TM$_{010}$ to different frequencies. Although this is equivalent to a rigid rod translation, this latter strategy is harder to realize for a cavity inserted in a cryogenics system.

\begin{figure}[!ht]
\begin{center}
	\includegraphics[width=1\linewidth]{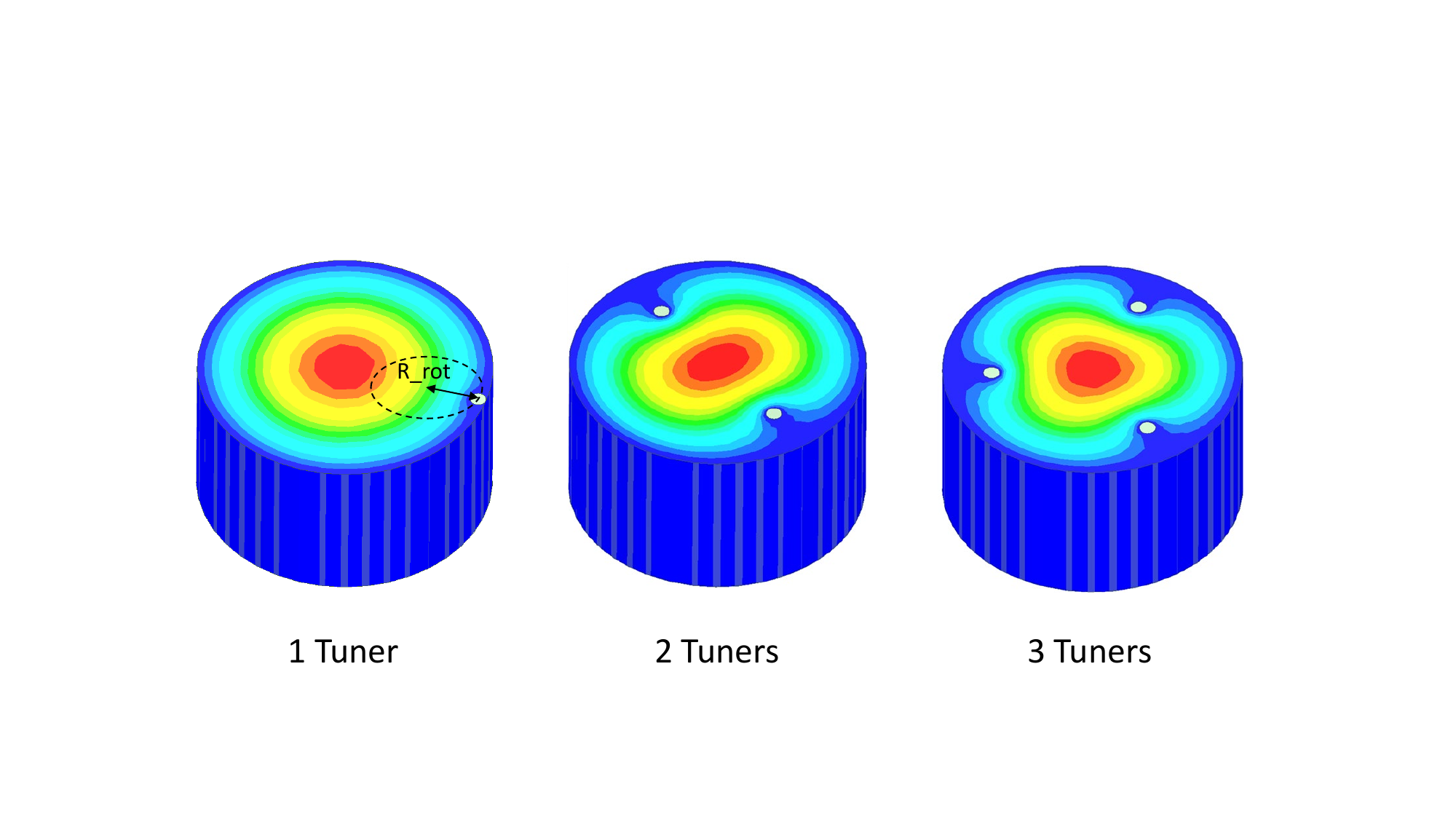}
	\caption{Sketch of the tuning rods configurations analyzed for the setup. Colors code the magnitude of the electric field strength.}
	\label{fig:field_tm010}
\end{center}
\end{figure}

\begin{table}[!ht]
  \begin{center}
  \vspace*{0.5cm}
    \begin{tabular}{c|c|c}
      Parameter& FLASH LF & FLASH HF ($S$)\\
      \hline\hline
      $R$ cavity (mm)               & 1050  & 590 \\ \hline
      $R_\mathrm{rod}$ (mm)  &   115  & 60 \\ \hline
      $R_\mathrm{rot}$ (mm)   &  276   & 160  \\ \hline
      $C_\mathrm{rot}$ (mm)   &   654 & 367 \\ \hline
      $L_\mathrm{rod}$ (mm)   &  1200 & 1200  \\ \hline
      n.\ tuning rods                    &   3      &    3  \\ \hline
      Frequency range (MHz)   & 117 - 206 & 206 - 360 \\ \hline
      $Q/1000$                         & 570 - 450 & 524 - 380 \\ \hline
      Form factor                      & 0.63 - 0.73 & 0.64 - 0.75 \\ \hline
      BW (Hz) @ $\beta = 1$   &   410 - 916 & 786  - 1895  \\ \hline\hline
    \end{tabular}
  \caption{Main parameters of the two cavities with frequency tuning system.}
  \label{tab:param}
  \end{center}
\end{table}
The contact between the rod and the parallel plates of the cavity must be guaranteed to avoid a deterioration of the quality factor. In fact, a non-perfect contact increases the losses by creating a gap that acts as a capacitor with spurious modes. Simulations show that a gap 1\,mm wide leads to a deterioration of the quality factor by less than 5\%. This problem was already addressed in Ref.~\cite{Lee:2015aht}. The different rod-configurations we explored are schematically shown in Fig.~\ref{fig:field_tm010}. 

During the preliminary assessment of the schematics, we have considered various configurations consisting of one, two, or three symmetric rods, with different radii, positions, and centres of rotation. Based on the calculation of the frequency range, the quality factor and the form factor of the working mode in all these configurations, a configuration consisting of three rods has been adopted. This choice, along with the size of the rods of radius R$_{\rm Large\, Rod} = 115$ mm for the larger cavity
and R$_{\rm Small\, Rod} = 60$ mm for the smaller cavity,
compromise between the complexity and tunability of the system. The frequency range and the number of required frequency steps for the two foreseen runs with the different resonant cavities are summarized in Table~\ref{tab:tuning}.
\begin{table}[!ht]
  \begin{center}
    \caption{Summary of frequency tuning ranges and steps.}
    \label{tab:tuning}
    \begin{tabular}{c|c|c}
      Stage &  Frequency range & Number of steps \\ \hline
      FLASH LF & (117 - 206)\,MHz & 80\,600 \\
      FLASH HF & (206 - 360)\,MHz & 69\,000 \\
      \hline\hline
    \end{tabular}
  \end{center}
\end{table}

\begin{figure}[!ht]
\begin{center}
	\includegraphics[width=1\linewidth]{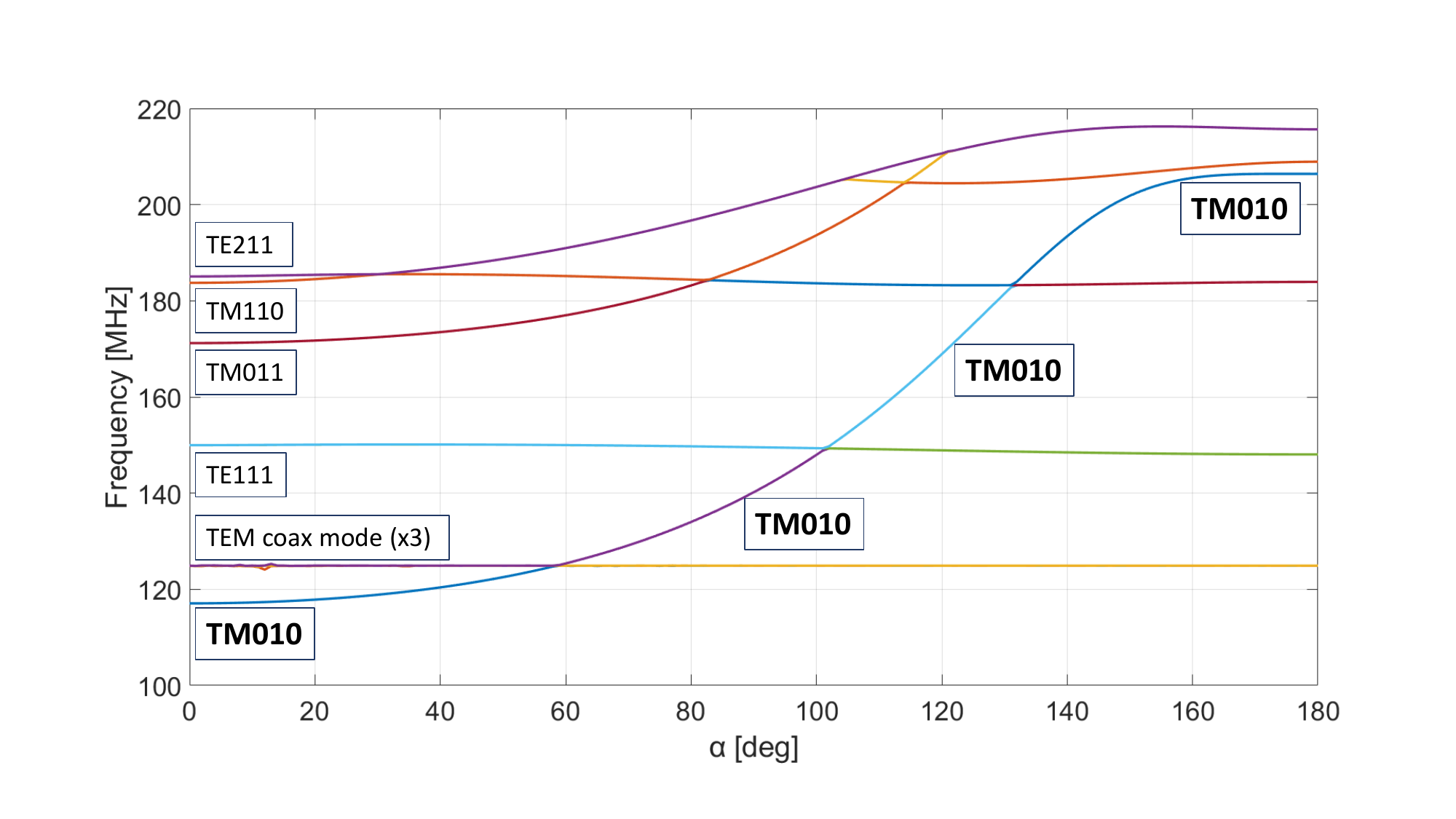}
	\caption{Complete mode mapping in the case of cavity with larger radius (Low Frequency).}
	\label{fig:tm010_lf}
\end{center}
\end{figure}

In Fig.~\ref{fig:tm010_lf} we show the cavity frequency as a function of the rotation angle $\alpha$ of the TM$_{010}$ and of the other identified modes for the three-rods case. Here, the three rods are moving symmetrically. The rotation angle $\alpha = 0$° corresponds the rod near the cavity wall. We note that this tuning system does not affect the mode frequencies equally, leading, for some rod positions, to an overlap between different modes. We note, for instance, the overlap of the working mode with TE modes and with the TEM mode generated in the the coaxial line formed by the rod (inner conductor) and the outer cavity wall (outer conductor). These are labelled  as ``TEM coaxial modes'' in the picture. Modes overlap lead to a ``mode mixing'' that makes the detection of the mode TM$_{010}$ more difficult in the crossing region~\cite{Lee:2015aht}.

\begin{figure}[!ht]
\begin{center}
	\includegraphics[width=0.6\linewidth]{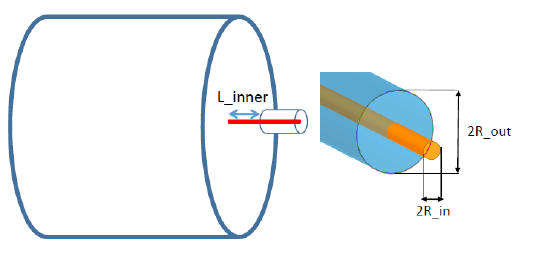}
	\caption{Sketch of the antenna coupled to the cavity modes.}
	\label{fig:ant}
\end{center}
\end{figure}

\begin{figure}[!ht]
\begin{center}
	\includegraphics[width=1\linewidth]{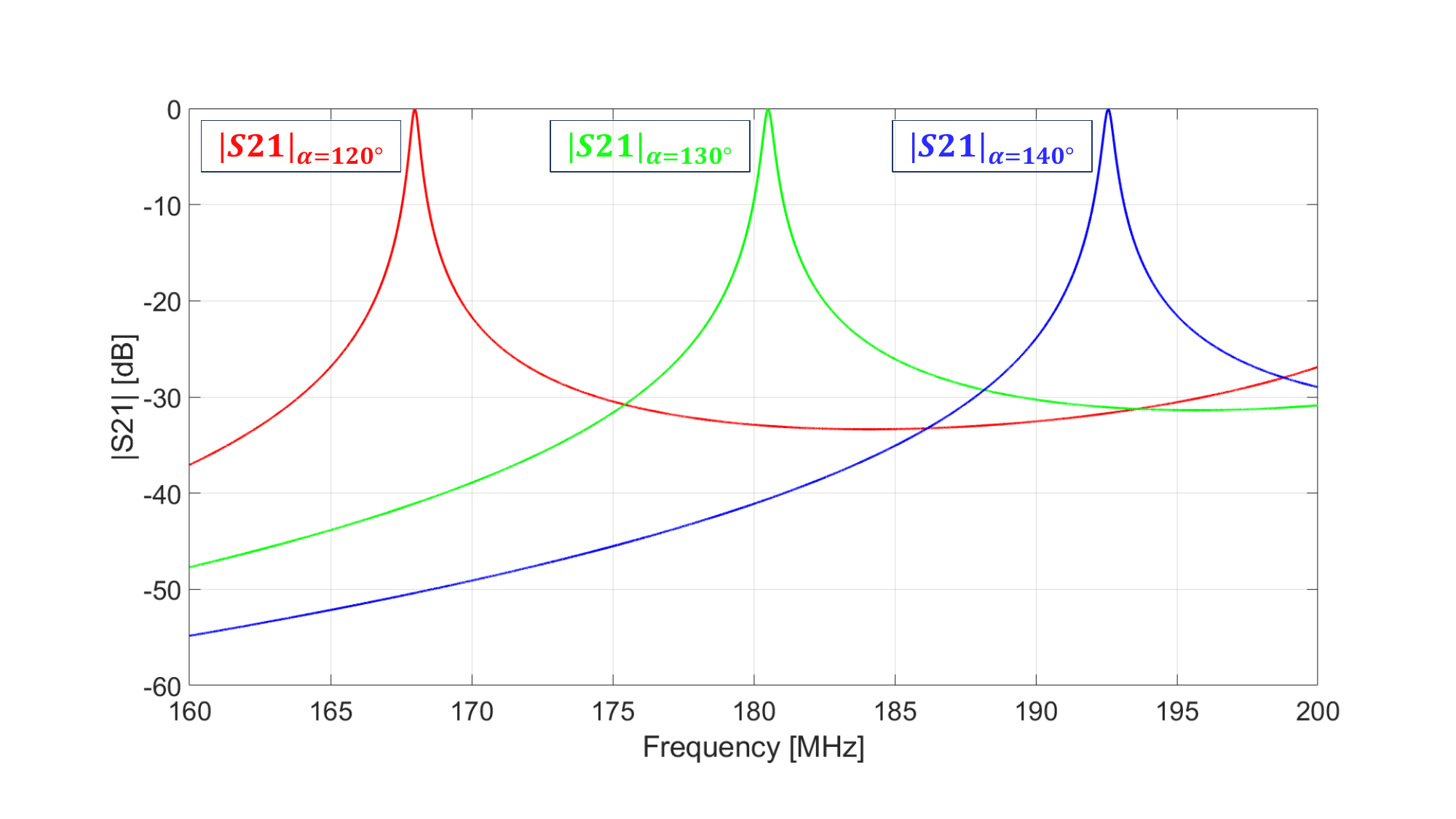}
	\caption{Transmission coefficient between two coaxial probes coupled to the cavity for different tuner positions in the frequency region of the TM$_{010}$ - TE$_{211}$ mode-mixing.}
	\label{fig:trans}
\end{center}
\end{figure}

The signal from the TM$_{010}$ mode is extracted through a coaxial probe inserted in one of the two end-caps of the cavity, parallel to the axis of the cavity itself, as sketched in Fig.~\ref{fig:ant}.
The probe coupling $\beta$ will be varied by changing the penetration depth of the inner conductor ($L_\mathrm{inner}$). Simulations show that a change in the penetration depth of a standard-SMA antenna by few tens of cm the coupling $\beta$ varies in a wide range between 0 and 2. We also verified that the TM$_{010}$ mode is always  well identifiable, in particular near the crossing region between TE-TM modes. This is due to the fact that the probe only couples to the longitudinal electric field of the mode, which is equal to zero for ideal TE and TEM modes and leads to a negligible coupling between the TM$_{010}$ mode and TE modes. This test was performed by simulating the transmission coefficient between two coaxial probes coupled to the cavity for different rods positions such that the frequency is in the TM$_{010}$-TE$_{211}$ mode-mixing region, see Fig.~\ref{fig:trans}. The result was also confirmed by simulations in a smaller frequency range. Residual mode crossings will be resolved by proper rotating the three tuning rods or by the proper insertion of small dielectric tuning rods.

\begin{figure}[!ht]
\begin{center}
	\includegraphics[width=0.8\linewidth]{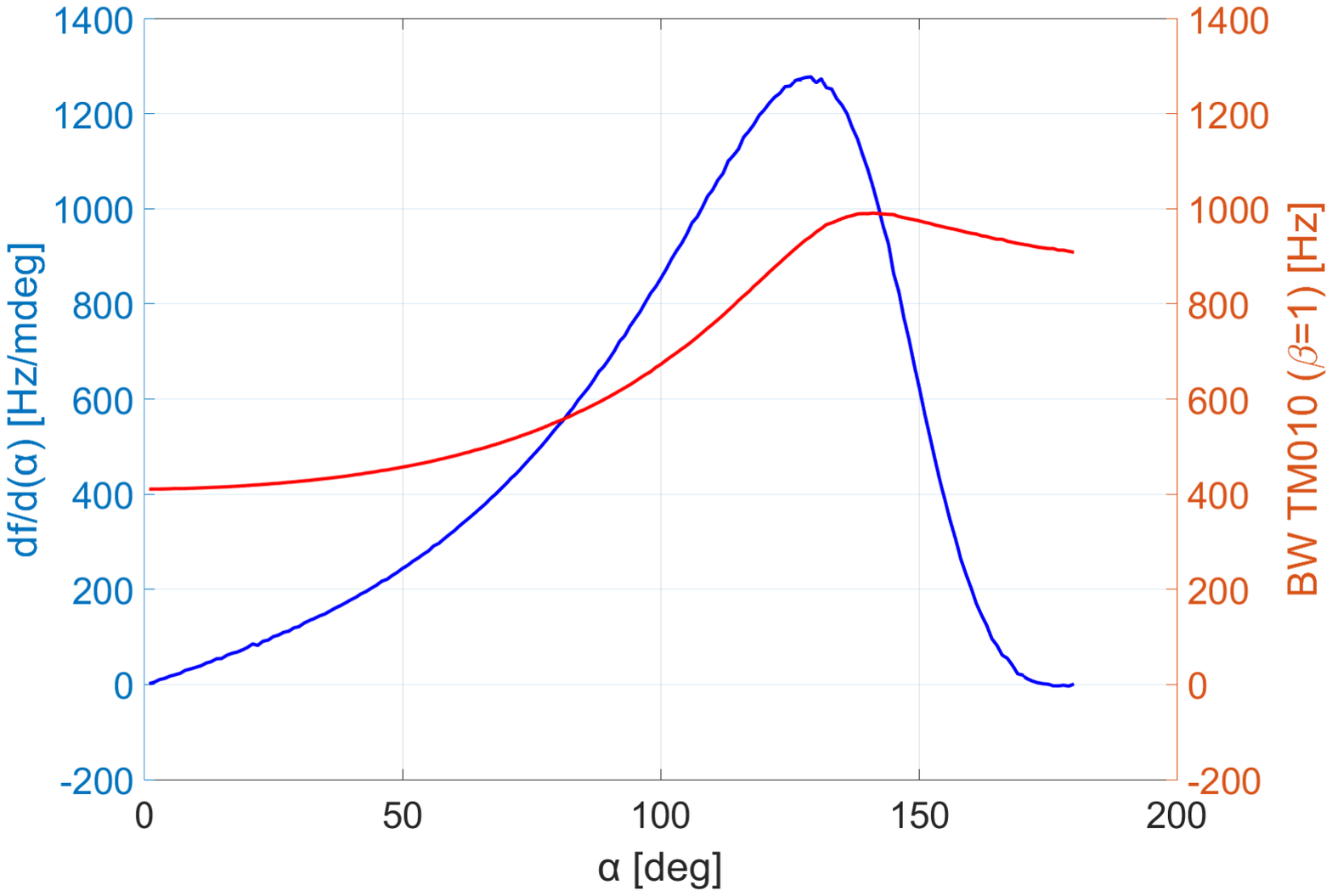}
	\caption{Sensitivity of the resonant frequency and TM$_{010}$ mode bandwidth to the rotation angle.}
	\label{fig:sens}
\end{center}
\end{figure}

\begin{figure}[!ht]
\begin{center}
    \hspace*{-1.0cm} 
    \includegraphics[width=0.45\linewidth]{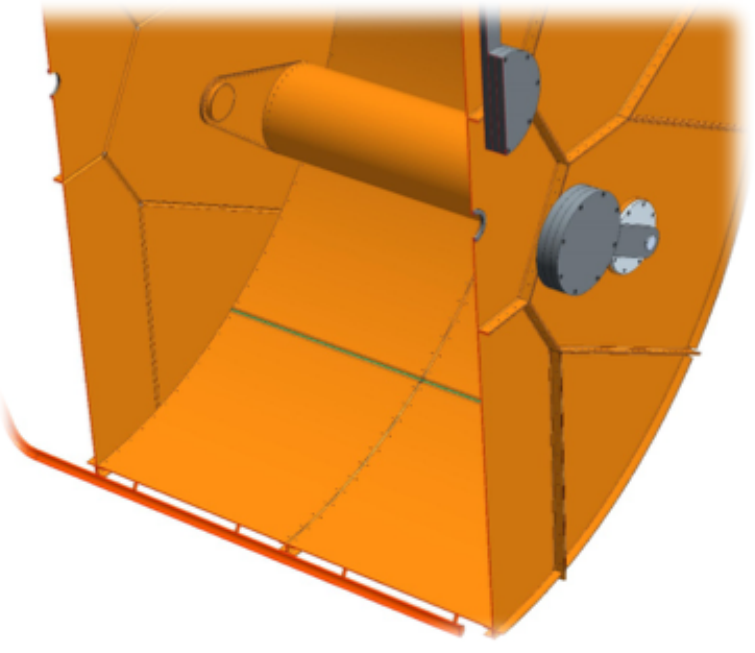}
    \hspace*{1.0cm}
	\includegraphics[width=0.45\linewidth]{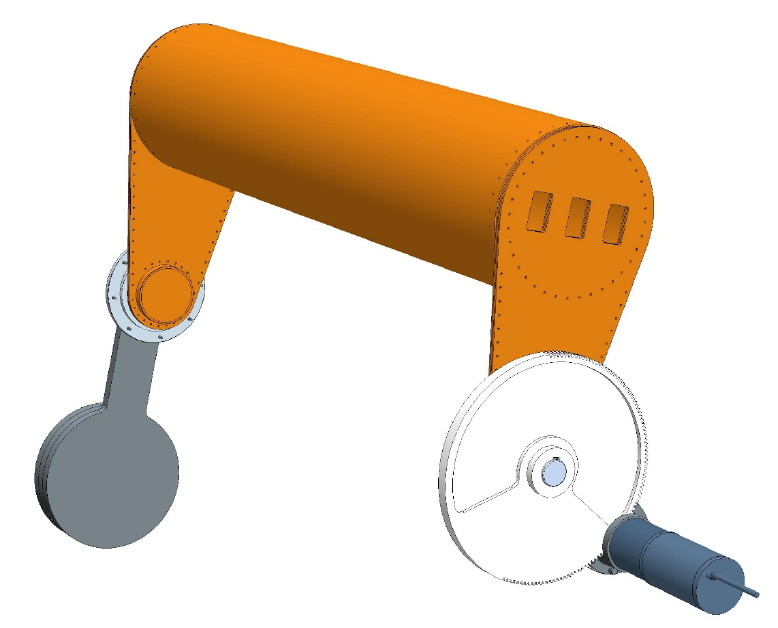}
	\caption{Internal view of the cavity with the tuner (left)~\cite{Alesini:2019nzq}; tuner assembly (right).}
	\label{fig:tuner}
\end{center}
\end{figure}

The minimum angle of rotation achievable with the FLASH tuning system must correspond to a frequency variation on the order of the cavity bandwidth reported in the last line of Table~\ref{tab:param}. The sensitivity of the resonant frequency to the rotation angle is shown in Fig.~\ref{fig:sens}.
From this figure, we deduce that the minimum tuning angle of rotation must be less than 12 $\mu$rad in the positions range from 80 degrees to 140 degrees. This will be assured by equipping each tuner by a vacuum compliant Hybrid Stepper-Motor (200 step/turn, 150\,min$^{-1}$) with a three-stage epicyclic gear unit (1:200) and a gear couple (1:10). The total angular step will be $9^{-4}$ degrees ($\sim2.5\,\mu$rad) (Fig.~\ref{fig:tuner}). The whole turn will take around 7 min.

\begin{figure}[!ht]
\begin{center}
	\includegraphics[width=0.8\linewidth]{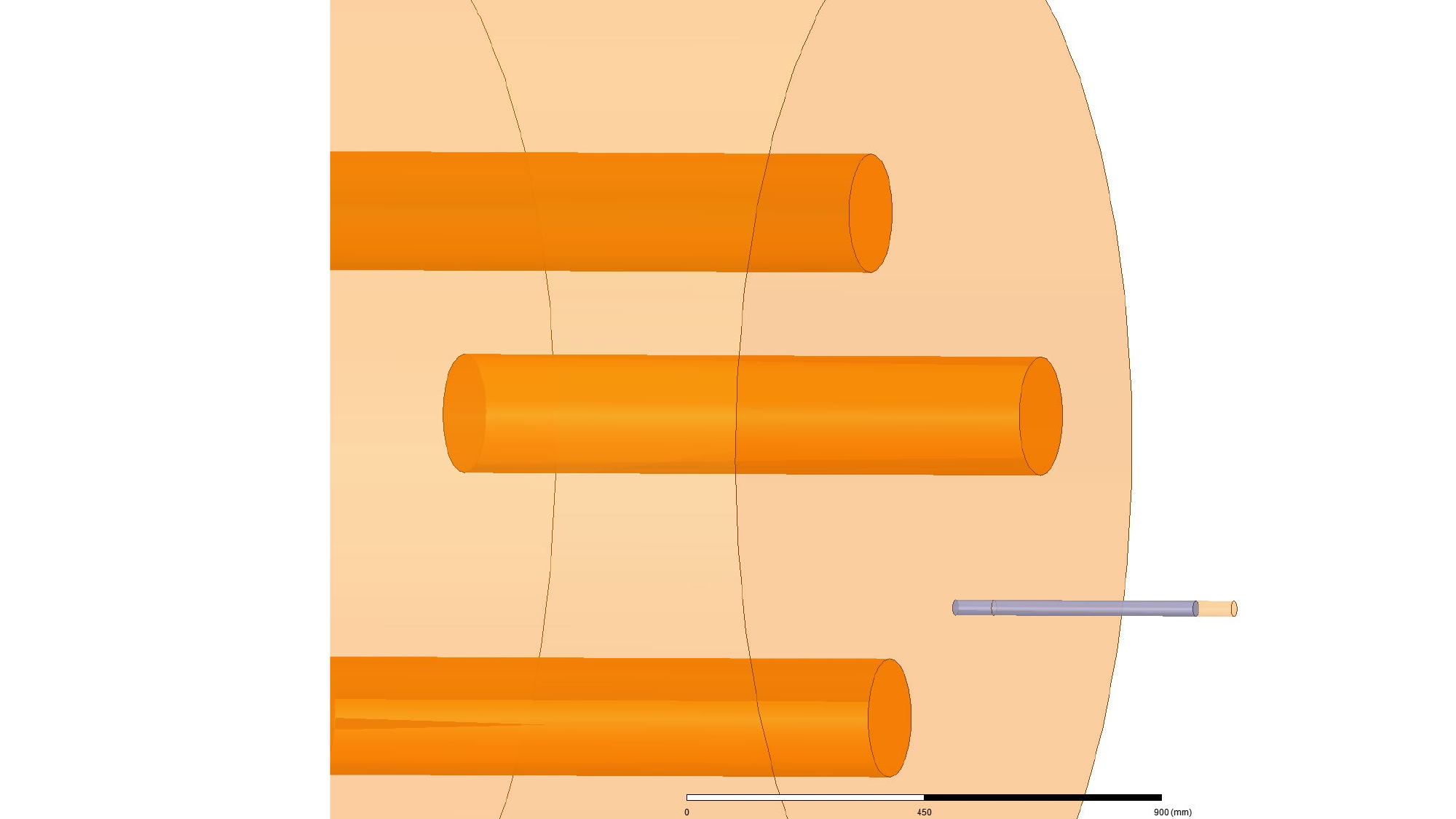}
	\caption{Dielectric tuning system for fine frequency tuning.}
	\label{fig:diele}
\end{center}
\end{figure}

To achieve a more precise tuning a small cylinder of dielectric material (alumina or sapphire) can be inserted as sketched in Fig.~\ref{fig:diele}. With a cylinder of 10 mm of radius, in sapphire, we achieved in the simulation a sensitivity of 160\,Hz\,mm$^{-1}$ that can be reduced or increased by decreasing or increasing the diameter of the cylinder itself, or by changing its position on the cavity.

\begin{figure}[!ht]
\begin{center}
	\includegraphics[width=1.0\linewidth]{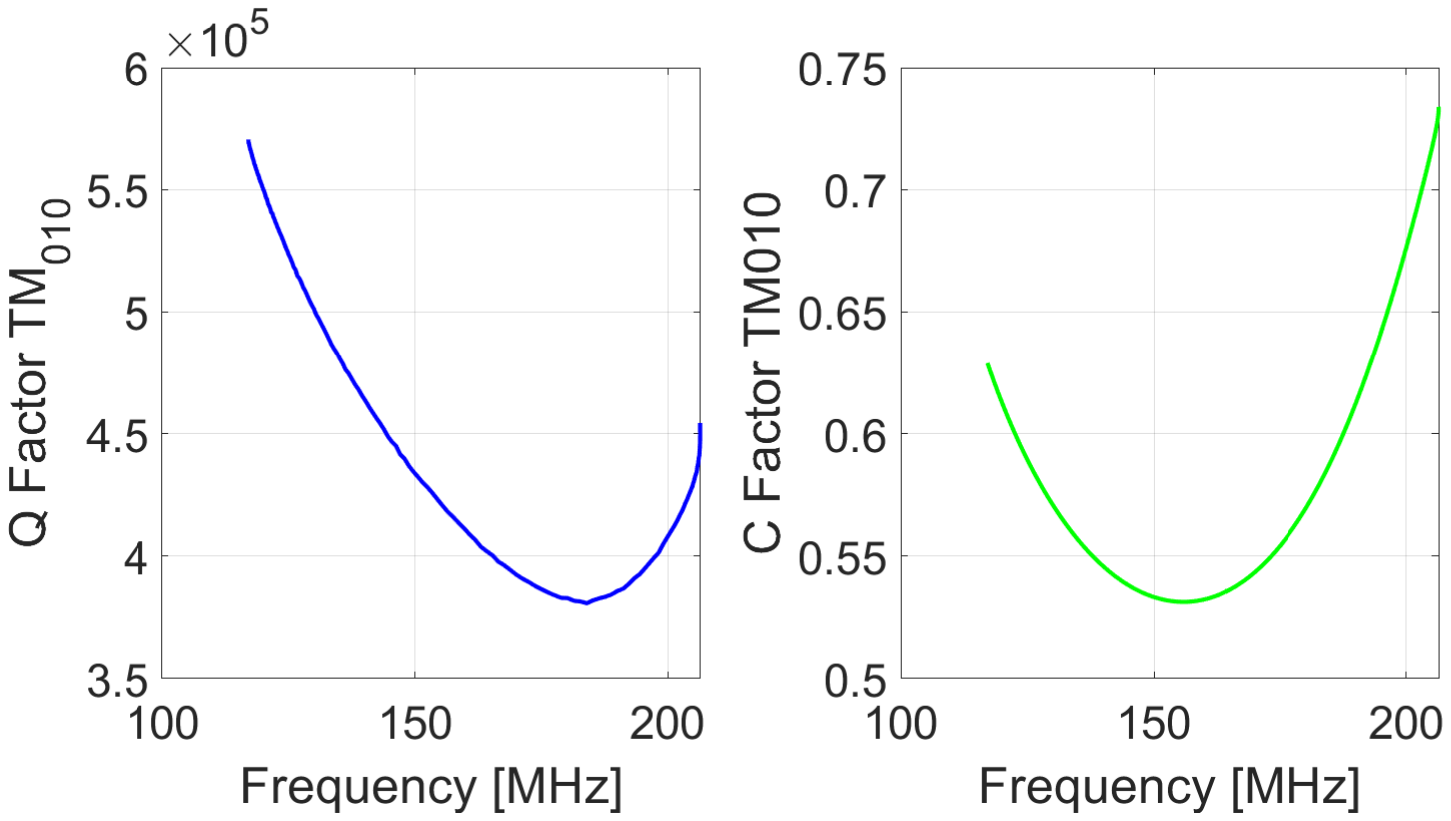}
	\caption{Quality (left) and Form (right) factors of the TM$_{010}$ mode as a function of frequency for Low Frequency cavity.}
	\label{fig:qual}
\end{center}
\end{figure}

In Fig.~\ref{fig:qual} we show the quality factor and form factor of the TM$_{010}$ mode as a function of frequency, where we assumed that the rods and the cavity
are made of copper with a value of RRR equal to 50 and, as a consequence, with a conductivity at 4\,K equal to $\sigma = 2.9 \times 10^9\,$S/m~\cite{Reuter:1948}. 

\begin{figure}[!ht]
\begin{center}
	\includegraphics[width=0.7\linewidth]{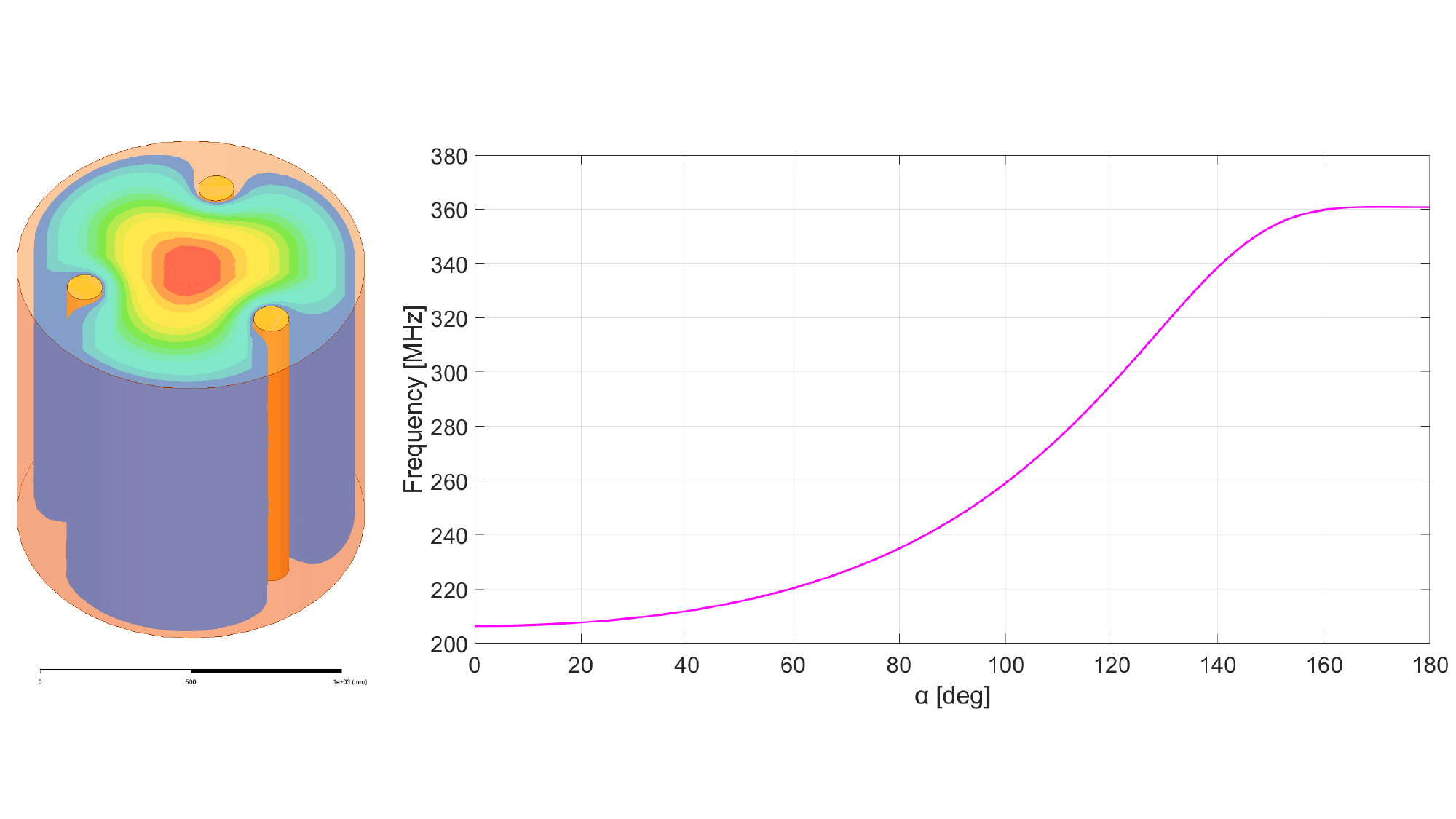}
	\caption{Cavity with smaller radius with tuning system (left); frequency of the
	TM$_{010}$ mode as a function of the tuner position (right).}
	\label{fig:cav2}
\end{center}
\end{figure}

\begin{figure}[!ht]
\begin{center}
	\includegraphics[width=0.8\linewidth]{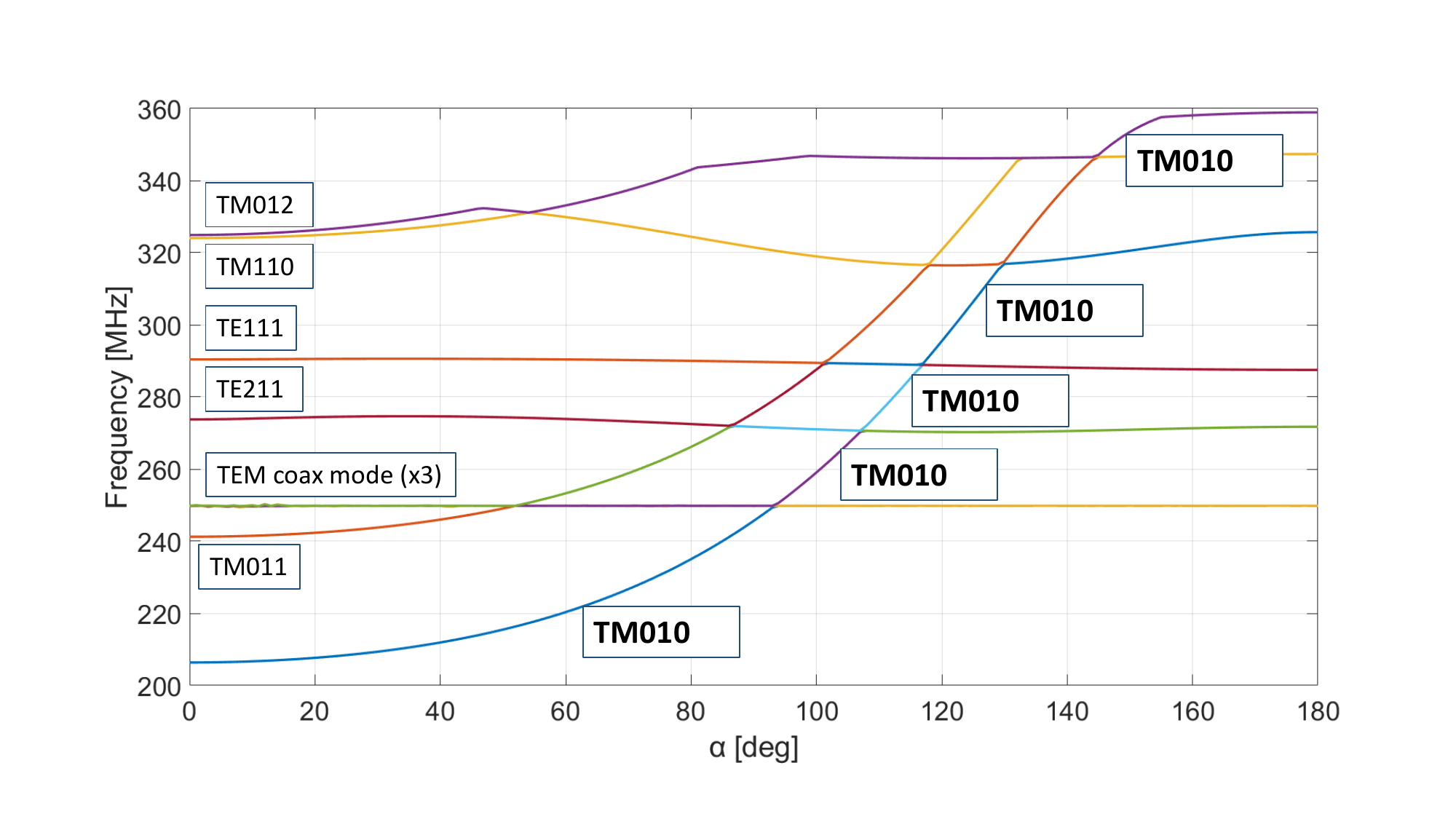}
	\caption{Complete mode mapping in the case of cavity with smaller radius (High Frequency).}
	\label{fig:tm010_hf}
\end{center}
\end{figure}

\begin{figure}[!ht]
\begin{center}
	\includegraphics[width=0.8\linewidth]{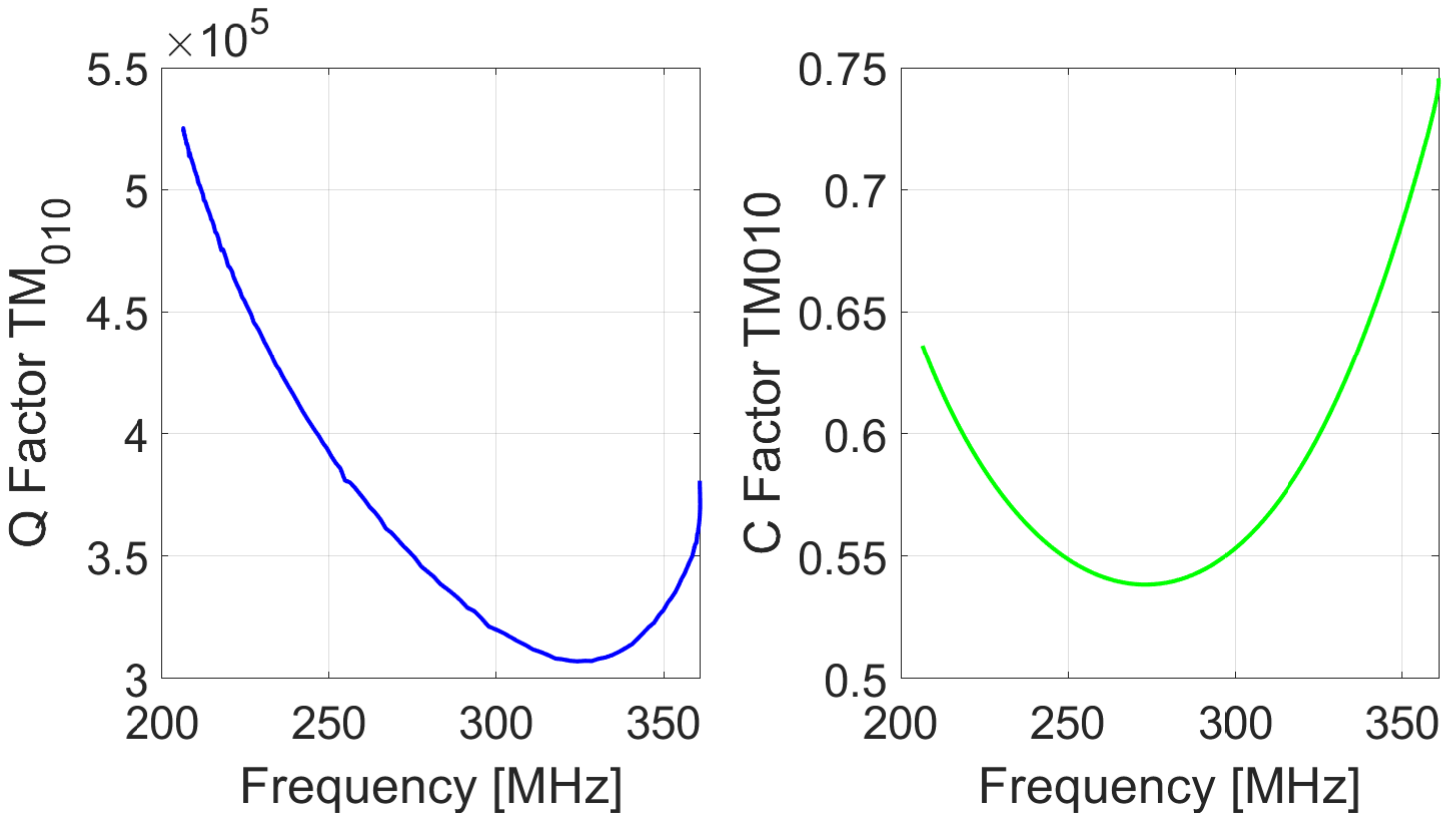}
	\caption{Quality (left) and form (right) factors as a function of frequency.}
	\label{fig:quff3}
\end{center}
\end{figure}

The three tuning rods of the smaller cavity, used to probe the higher frequency range, have radius equal to 60 mm. The cavity with the rods is schematically shown in the left panel of Fig.~\ref{fig:cav2}.
In the right panel we show the frequency of the TM$_{010}$ mode as a function of the tuner position. The complete mode mapping is given in Fig.~\ref{fig:tm010_hf} while the quality factor and the form factor are shown in Fig.~\ref{fig:quff3}.

\begin{figure}[!ht]
\begin{center}
	\includegraphics[width=0.8\linewidth]{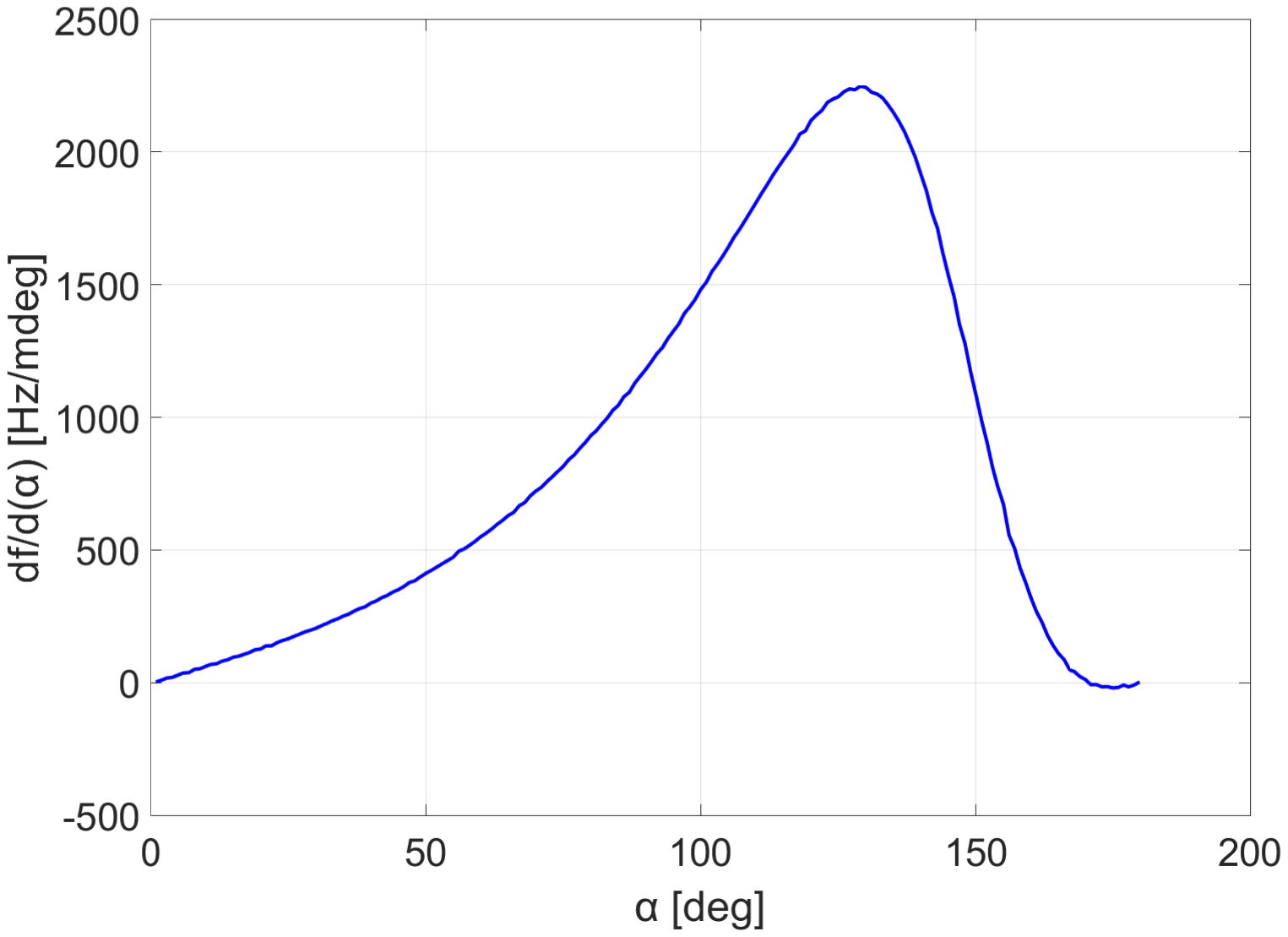}
	\caption{Sensitivity of the tuning system for the cavity with smaller radius.}
	\label{fig:sens2}
\end{center}
\end{figure}

Finally, the sensitivity of the rod tuning-system for the small cavity is reported in Fig.~\ref{fig:sens2}. 

\subsection{Other modes relevant for axion, chameleon and HFGW searches}

In order to exploit the full discovery potential of FLASH we must take into account modes higher than the TM$_{010}$, that couple to the dark-matter axions but also to dilatons, chameleons and high frequency GWs (HFGW). The signal from the TM$_{020}$ mode can be acquired to increase the frequency range of the axion and dark-photon search. For this mode the coupling factor is $C_{020}=0.13$, about a factor 5 smaller than for the TM$_{010}$ mode. The mode TE$_{011}$ is relevant for chameleons searches as discussed in section~\ref{ch:sec_chameleons}. For HFGW both TE and TM modes are relevant and simultaneous detection of the different modes may be important to increase the signal sensitivity and to exploit the detector directionality~\cite{Berlin:2022hfx}. We report frequency and quality factor for some of these modes in Tab.~\ref{tab:temodes} and~\ref{tab:tmmodes}.
\begin{table}[!ht]
  \begin{center}
  \vspace*{0.5cm}
    \begin{tabular}{c|c|c}
      Mode & Frequency (MHz) & Q$_0$ factor (at 4~K)\\
      \hline\hline
        TE$_{111}$ & 150.4  & $711\times 10^3$ \\ \hline
        TE$_{112}$ & 263.5  & $871\times 10^3$ \\ \hline
        TE$_{211}$ & 186.9  & $735\times 10^3$ \\ \hline
        TE$_{212}$ & 285.9  & $817\times 10^3$ \\ \hline
        TE$_{011}$ & 214.5  & $1.3\times 10^6$ \\ \hline\hline
    \end{tabular}
  \caption{TE Modes for the Large cavity}
  \label{tab:temodes}
  \end{center}
\end{table}

\begin{table}[!ht]
  \begin{center}
  \vspace*{0.5cm}
    \begin{tabular}{c|c|c}
      Mode & Frequency (MHz) & Q$_0$ factor (at 4~K)\\
      \hline\hline
        TM$_{010}$ & 109.5  & $626\times 10^3$ \\ \hline
        TM$_{011}$ & 166.1  & $526\times 10^3$ \\ \hline        
        TM$_{110}$ & 174.4  & $790\times 10^3$ \\ \hline
        TM$_{020}$ & 251.2  & $948\times 10^3$ \\ \hline 
        TM$_{111}$ & 214.5  & $598\times 10^3$ \\ \hline
        TM$_{012}$ & 272.3  & $752\times 10^3$ \\ \hline
        TM$_{112}$ & 304.7  & $712\times 10^3$ \\ \hline
        TM$_{210}$ & 233.7  & $915\times 10^3$ \\ \hline
        TM$_{211}$ & 264.9  & $664\times 10^3$ \\ \hline
        TM$_{212}$ & 342.1  & $755\times 10^3$\\ \hline\hline
    \end{tabular}
  \caption{TM Modes for the Large cavity}
  \label{tab:tmmodes}
  \end{center}
\end{table}

\begin{figure}[!ht]
\begin{center}
	\includegraphics[width=1.0\linewidth]{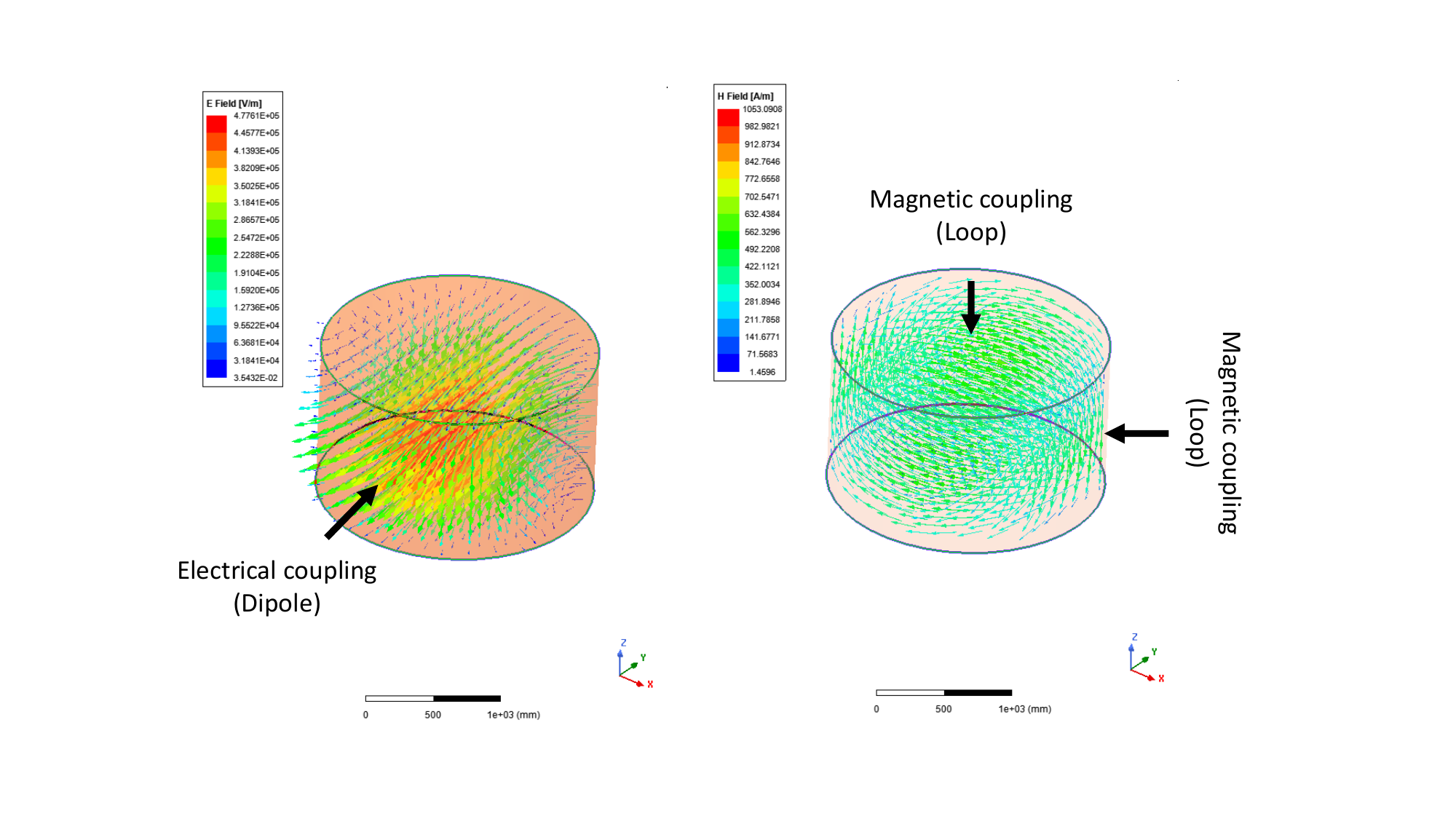}
	\caption{Electric field (left) and magnetic field (right) of TE$_{111}$ mode with a sketch of the insertion points for the two different types of coupling.}
	\label{fig:TE111_coupling}
\end{center}
\end{figure}

While the signal from the TM modes can be acquired from the same dipole antenna discussed in the previous section, the TE modes require a different kind of antenna and/or insertion point. In fact, also the TE modes can be acquired by means of a coaxial dipole, placed however in a different insertion point with respect to the TM modes. In general for cylindrical cavities this region is on the lateral surface. This type of signal extraction is called electrical coupling.
Signals from TE modes can also be extracted by magnetic coupling created by bending the inner conductor of the coaxial antenna into a loop. This antenna is placed in an area of the cavity where the magnetic field is maximum and transversal to the surface of the loop. The coupling of the $\beta$ probe will be varied by changing the angle between the loop surface and the magnetic field, given the loop size.
Fig.~\ref{fig:TE111_coupling} shows the electric field and magnetic field of TE$_{111}$ mode with an example of electric coupling and magnetic coupling and their insertion point.
A further comment is required for TE modes. The resonant cavity will be composed by tiles, realized by cold formed commercial plates of copper OFHC joint by rivets~\cite{Alesini:2019nzq}. To minimize eddy current loops, in particular in case of a magnet quench, we planned to insulate the longitudinal joint between tiles by NEMA G10 plates and insulating sleeves. This will however interrupt the currents of TE-modes excitations inhibiting it, requiring a different solution to be used.

\section{The FLASH cryogenics}
\label{sec:cryogenics}

The RF cavity for FLASH must be cooled at low temperature to reach the needed sensitivity. To do so, it will be hosted in a dedicated custom cryostat and cooled at 4.5\,K using liquid helium (LHe). The cryostat will be composed by an external stainless steel vacuum vessel, containing an aluminum-alloy radiation shield kept at about 70\,K by cold gaseous helium (GHe) and surrounding the cavity. Both the cavity and the shield are cooled in contact with pipes in which the helium flows.

The FLASH experiment takes advantage of the availability at LNF of two existing apparatuses: the DA$\Phi$NE cryogenic plant~\cite{ligi:2002} and the FINUDA magnet. The cryogenic plant is a liquid helium refrigerator/liquefier, which kept cooled the KLOE and FINUDA superconducting magnets during their runs for about twenty years. The plant is capable to be connected and cool both the FINUDA magnet and the FLASH cryostat at the same time.
\begin{figure}[!ht]
\begin{center}
	\includegraphics[width=0.8\linewidth]{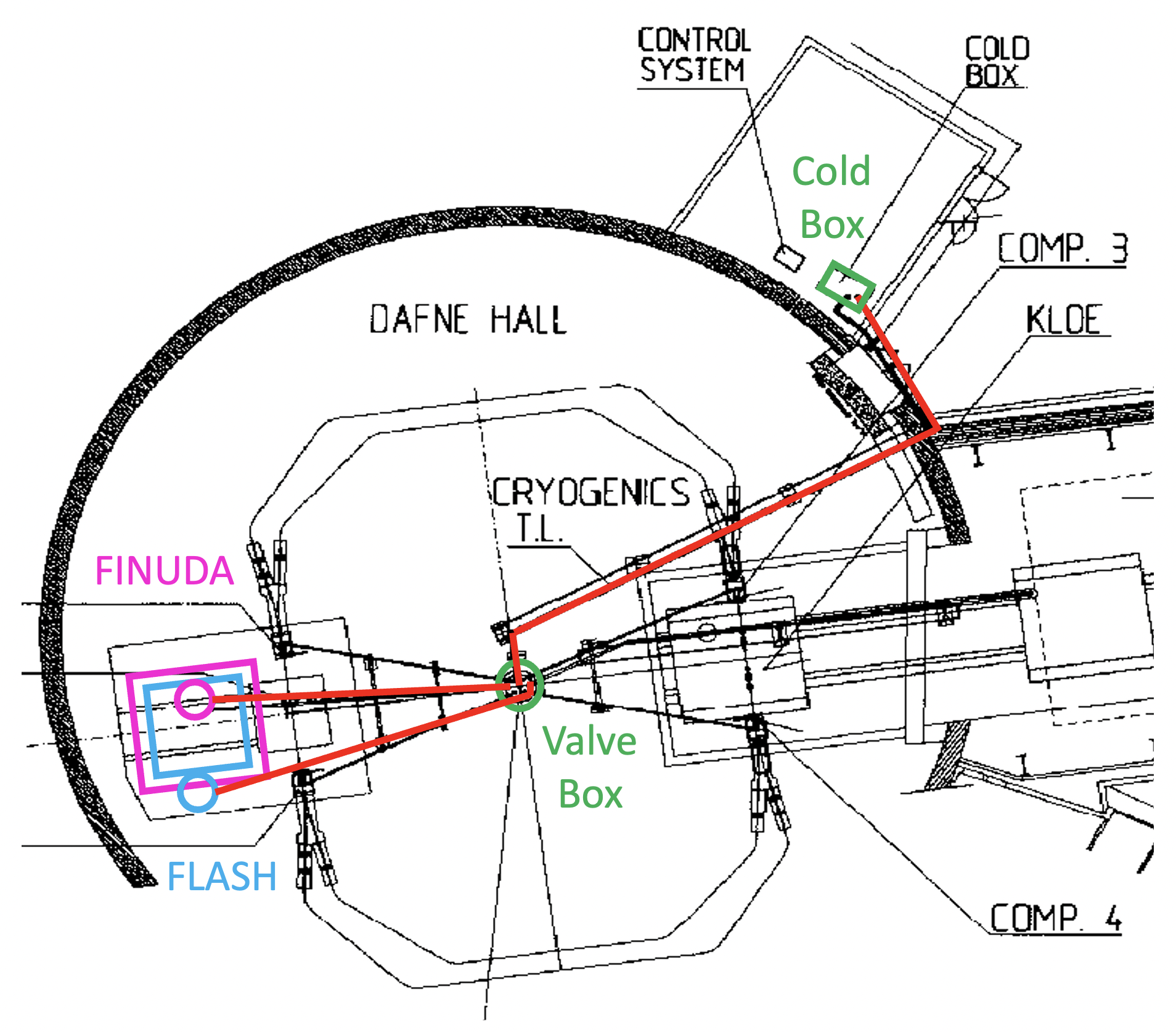}
	\caption{General layout of the FLASH cryogenics}
	\label{fig:cryolayout}
\end{center}
\end{figure}

The general layout of the FLASH cryogenics is shown in Fig.~\ref{fig:cryolayout}. The cryogenic plant, a LINDE TCF50 refrigerator/liquefier, is composed by a Cold Box, in which the gas is cooled down both to 5.2\,K/3\,bar (so in supercritical state, SHe) and to 70\,K/5\,bar, and a Valve Box, where the cold helium comes from the Cold Box and is distributed to the users. Cold and Valve Boxes are connected via a transfer line (in red in the figure).

The Valve Box has two main send/return connections for the users, formerly used for KLOE and FINUDA. Now, the FLASH cryostat (blue square in the figure) can be connected in place to the ex-KLOE line.

\subsection{The cryogenic plant}

The cryogenic plant is based on a LINDE TCF~50 Cold Box and a KAESER ESD442 compressor. As said, the refrigerator can provide both SHe at 5.2\,K and GHe at about 70\,K, the latter for the radiation shields cooling. Both cryogens are then sent in a distribution box (Valve Box) where they are split to the users. The last cooling from 5.2\,K to 4.5\,K is carried out directly inside the users, in dedicated service turrets equipped with a J-T valve, for the last helium expansion/liquefaction, and a buffer volume, for the liquid storage.

The overall cooling capacity of the plant is as following:
\begin{itemize}
  \item liquefaction rate = 1.14~g/s;
  \item Refrigeration capacity at 4.45~K/1.22~bar = 99~W;
  \item Shield cooling capacity below 80~K~=~800~W.
\end{itemize}

\subsection{The FINUDA superconducting magnet}

The FINUDA Magnet~\cite{bertani1999finuda} is an iron shielded superconducting solenoid coil, made by Ansaldo Energia (Italy). The coil is cooled with thermo-siphoning method, and its service turret has a buffer reservoir of about 25 liters. The magnet works in continuous cooling at the temperature of 4.5~K. Its nominal magnetic field is 1.1~T with a current of 2796~A and the stored energy is 10.34\,MJ. The magnet bore dimensions consist of a length of 2.4\,m and a diameter of 2.7\,m. It has operated until 2007 for the FINUDA (FIsica NUcleare at DA$\Phi$NE) experiment at INFN-LNF.\footnote{\href{http://www.lnf.infn.it/esperimenti/finuda/finuda.html}{http://www.lnf.infn.it/esperimenti/finuda/}}

\subsection{The FLASH cryostat}

The FLASH cryostat must maintain the RF cavity at the LHe temperature. It must withstand several requirements:
\begin{itemize}
  \item it should match the available space inside the FINUDA magnet bore, and should be provided by all the mechanical tools for its insertion on the bore.
  \item it should be supported either by the magnet itself or by the ground, using suitable legs,
  \item it should be supplied by the cryogenic plant, so it must be equipped with a cryogenic service turret,
  \item it should contain a 300\,mK small-size refrigerator for the low-T amplification device.
\end{itemize}
Most of the general considerations made for the KLASH cryostat~\cite{Alesini:2019nzq} are still valid for FLASH. 

\subsection{SQUID cooling at 300~mK}
In order to minimize the thermal noise generated on the SQUID, it is possible to lower its temperature using a compact-size $^3$He evaporation refrigerator, capable to cool it down to about 300~mK.

This kind of refrigerator is commercially available,\footnote{\href{https://www.chasecryogenics.com}{https://www.chasecryogenics.com}} and can be conveniently inserted inside the FLASH cryostat. It is actually a two-stage refrigerator, where a first stage is a $^4$He refrigerator, used to pre-cool at about 1\,K and allows the condensation of the the $^3$He present in the second stage. In this way temperatures close to 300\,mK can be reached, depending on the thermal input from the SQUID.

The refrigerator requires some dedicated space inside the FLASH cryostat, as the SQUID must be placed in the proximity of the cavity antenna.

\section{Signal acquisition}
\label{sec:signalacquisition}

\subsection{Cryogenic Amplifiers}
\label{sec:amplifiers}

\begin{figure}[!ht]
\begin{center}
	\includegraphics[width=0.8\linewidth]{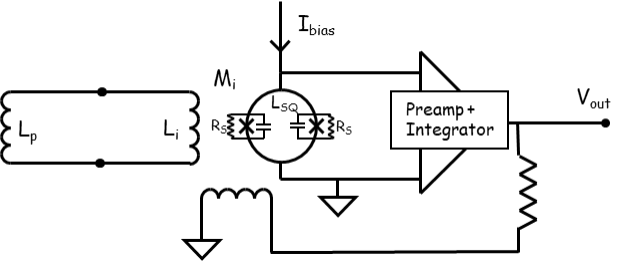}
	\caption{Flux locked DC Squid.}
	\label{fig:squid1}
\end{center}
\end{figure}

A Superconducting QUantum Interference Device (SQUID) is used as a sensitive magnetic flux detector in a variety of applications. It is also used as a low-noise, low-power-dissipation RF and microwave amplifier~\cite{doi:10.1063/1.121490}. At ultracryogenic temperatures, the noise scales with the temperature down to 200 - 300 mK. However, at lower temperatures, the Joule heating of the electrons gas due to the bias current causes the noise to deviate from the linear behavior and saturate~\cite{PhysRevB.75.104303}. This is due to the fact that the coupling between the electrons in the Josephson junction shunt-resistors $R_s$ (see Fig.~\ref{fig:squid1}) and the phonons is very weak.

Nevertheless, at frequencies below 1 GHz, this noise is an order of magnitude lower than the state-of-the-art semiconductor based amplifiers operating at low temperature. Commercially available two-stages dc-SQUID operating at 100 mK demonstrate an energy resolution of 30$\hbar$, a factor 30 above the quantum limit,
in the radio frequency band.
The use of dc-SQUID has been demonstrated in a flux-locked loop scheme~\cite{1439753}, in which the coil is directly inserted in the resonant cavity to pick-up the magnetic field, up to 130\,MHz. The operating frequencies of FLASH are slightly higher (from about 117\,MHz to about 360\,MHz). In addition, micro-vibrations of the apparatus in the presence of an high static field in the cavity, could induce spurious signals in the dc-SQUID. For this reason, a different detection scheme will be used.

A possible solution is the one used by ADMX~\cite{ADMX:2018gho} in which a Microstrip SQUID Amplifier (MSA)~\cite{Muck:1999ts, Muck_2010} is used as first amplifier and a cryogenic heterojunction field-effect transistor (HFET) amplifier is in the 4 K region. The MSA is an effective solution of the problem, allowing to operate with a gain depending on the frequency that is related to the microstrip length, while the bandwidth is defined by the microstrip impedance. 
\begin{figure}[!ht]
\begin{center}
	\includegraphics[width=0.7\linewidth]{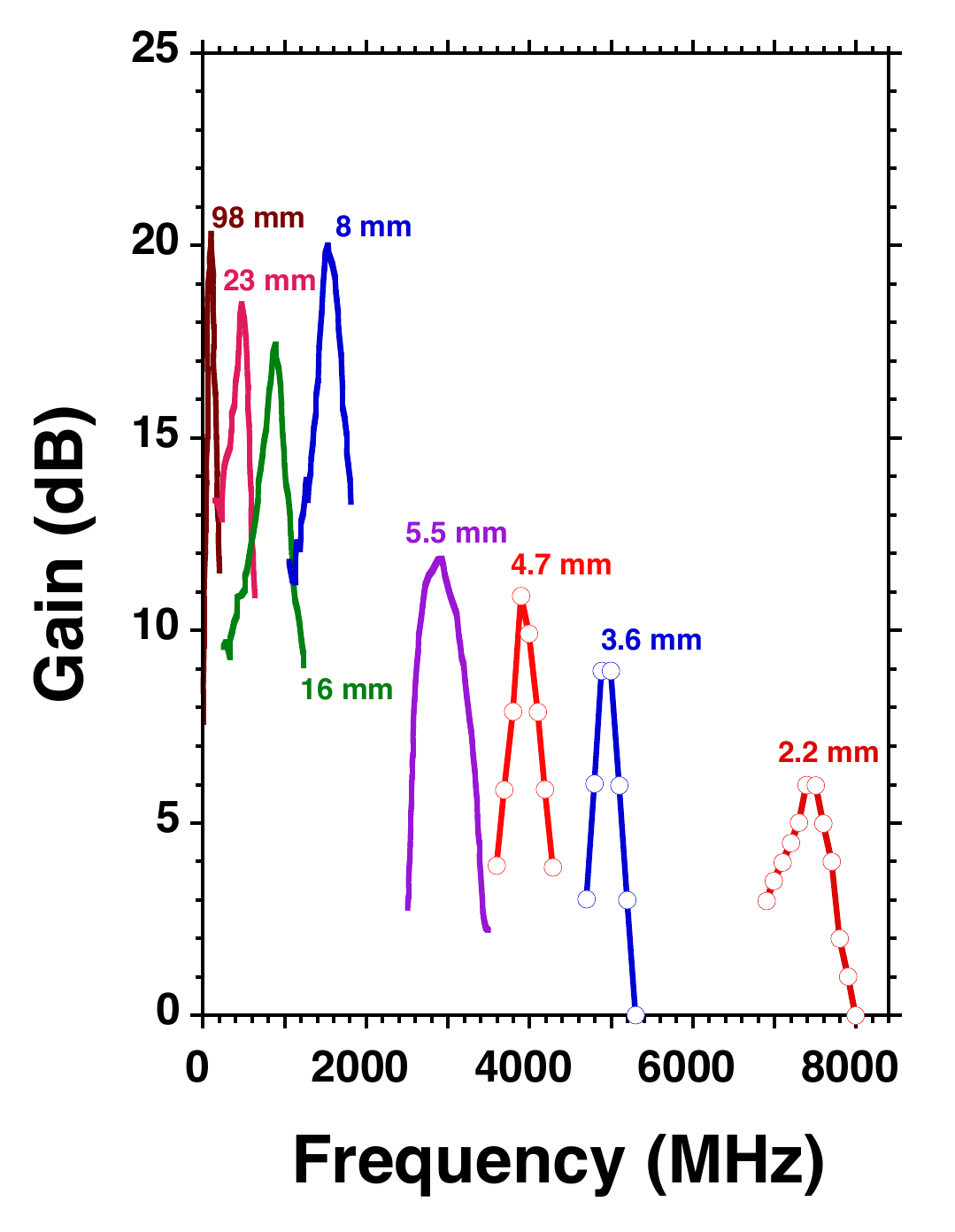}
	\caption{Gain vs frequency for MSA amplifiers. For each curve the length of the microstrip resonator is indicated. Figure from Ref.~\cite{doi:10.1063/1.1572970}.}
	\label{fig:squid2}
\end{center}
\end{figure}

\begin{figure}[!ht]
\begin{center}
	\includegraphics[width=0.7\linewidth]{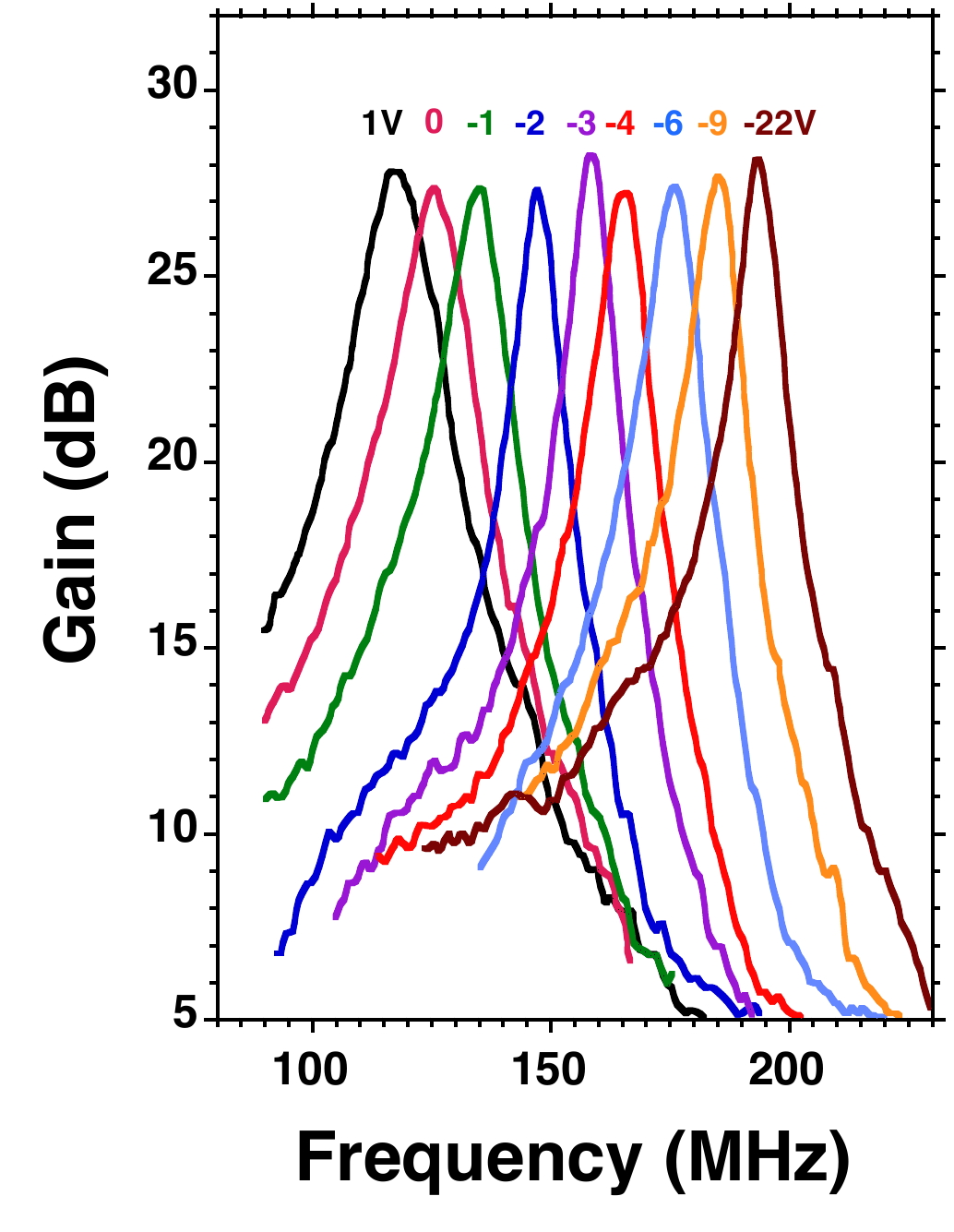}
	\caption{MSA gain vs frequency for different values of bias voltage applied to the varactor diode. Figure from Ref.~\cite{doi:10.1063/1.125383}.}
	\label{fig:squid3}
\end{center}
\end{figure}

For an operation frequency below 2 GHz, gains are over 20\,dB, see Fig.~\ref{fig:squid2} from Ref.~\cite{doi:10.1063/1.1572970}. To cover completely the frequency band required by FLASH a varactor diode connected to the input coil can be used. The capacitance is adjusted by changing the reverse bias resulting in an "effective" variation of the microstrip length. In Fig.~\ref{fig:squid3}, taken from Ref.~\cite{doi:10.1063/1.125383}, the gain is shown as a function of frequency for nine different values of the reverse bias applied to the varactor diode. 

The noise temperature $T_n$ of an MSA scales linearly with the operation frequency and the bath temperature \cite{doi:10.1063/1.1347384}. In Ref.~\cite{doi:10.1063/1.125383} a noise temperature of 170\,mK was observed with a bath temperature of 300\,mK and an operation frequency of 520\,MHz. We scale this value to the FLASH frequencies (ranging from a minimum of 117\,MHz for the large cavity, to a maximum of 360\,MHz for the small cavity) obtaining a noise temperature ranging from 38\,mK to 118\,mK (about a factor 7 with respect to the quantum limit in both cases).  

Both dc-SQUID and MSA are sensitive to magnetic field, mechanical vibrations and pick up noise. Adequate shielding must be used in order to decrease the static magnetic field in the cavity by 6 or 7 order of magnitude. We studied several possible scheme to shield the SQUID and to guarantee the correct thermalization. A passive approach makes use of both high-permeability and superconducting material to screen effectively both the axial and the radial magnetic field component in a semi-infinite tube~\cite{doi:10.1063/1.1141503}. An alternative solution is based on the use of a superconducting magnet in a persistent condition (for example NbTi) cooled in a zero field condition (ZFC). The last solution foresee the use of an active shielding using Helmholtz coils to generate an opposite magnetic field in the region where the SQUID is present. A magnetic filed sensor provides a negative feedback to drive the current in the cancelling coils. All these solutions were detailed in the conceptual design report of the KLASH experiment~\cite{Alesini:2019nzq}. 

\subsection{Room temperature amplification and data acquisition}

Thanks to the high gain in the cold section of the amplification chain ($>35$~dB), the room temperature amplification is not particularly demanding from the noise point of view. After a radio frequency (RF) amplification stage based on commercial low-noise components an intermediate frequency (IF) stage and a audio frequency (AF) mixer stage are foreseen. A power gain of between 30\,dB and 60\,dB may be required, and two amplification stages may be necessary, in order to match the dynamic range of the following stages.

An equivalent noise temperature up to 150\,K is acceptable for the RF amplifier(s). If the power gain of the previous stages is higher than 35\,dB, then the contribution of the first RF amplifier to the system noise temperature is lower than 0.05\,K and can be considered negligible. Low noise RF commercial amplifiers with a bandwidth of 500\,MHz are adequate for this purpose; for example, the aptmp3-00200005-1627-D9-lN-2 by AmpliTech, Inc.\ or the CITLF3 by Cosmic Microwave Technology, Inc., although an R\&D phase is still needed. The output of the RF stage feeds the IF stage organized with the AF stage in a double heterodyne receiver. A mixer, such as the IRM10-1000 by Sirius Microwave, with an insertion loss of 7\,dB and an image rejection better than 25\,dB is the first component of this stage. This mixer shifts the RF signal power from the tunable resonant frequencies of the cavity to a fixed intermediate frequency of 10.7\,MHz while rejecting image noise power. This image rejection is necessary to avoid mixing off-band RF power into the IF bandwidth, and the chosen value of the intermediate frequency is compatible with commercial devices.

The noise outside the signal bandwidth is reduced using a high selective bandpass filter with a bandwidth around few tens of kHz, compatible with the resonant cavity quality factor. In the AF stage the signal coming from the bandpass filter, after a further amplification stage to increase the signal to noise ratio, is down-converted to a near audio frequency band with a central frequency of 30\,kHz. The reference oscillator is obtained with a low phase noise frequency synthesizer with high frequency stability such the N5171 by Keysight. A commercial fast Fourier transform (FFT) spectrum analyser with a resolution bandwidth of hundreds hertz is used to acquire the power spectrum of the final AF mixer.

\section{Data Analysis}
\label{sec:dataanalysis}

\subsection{Axions}

The signal of axion conversion in haloscopes is detected as an excess in the the measured power spectrum. The analysis procedure relies in the acquisition of data by scanning the resonant frequency with a series of steps that cover a range of possible axion masses. For each value of the frequency scan, the power from axion-to-photon conversion is determined from the data, together with the measurement of experimental parameters that influence the power spectrum. 

The expected power produced by the axion-to-photon conversion is very small. Applying Eq.~\eqref{eq:power} to the FLASH experiment, the expected power released in the cavity is on the order of $P_{\rm sig} \simeq 10^{-22}\,$W. The analysis procedure must be designed to identify such a small signal in the presence of significant thermal and amplification-chain noise. A large number of power spectra must be collected for each resonant frequency and combined after subtracting the background contribution using a model of the acquisition chain. Ancillary data will be collected to measure the stability of the system and to characterize the noise components, with the help of a simulation model of the acquisition chain.

The data collection strategy for FLASH is based on the use of two cavities. Three metallic and one dielectric rods are used to tune the resonant frequency of the ${\rm TM}_{010}$ mode in both cavities. To speed up the scanning rate the antenna will be over-coupled with coupling $\beta=2$. For each position of the tuning rods the antenna, gain and coupling will be measured and optimized, the resonance frequency and the quality factor of the cavity determined and finally a given number of power spectra acquired. After these measurements the procedure will be repeated with a new value for the resonant frequency obtained by changing the position of the rods. 

The importance of measuring the unloaded quality factor $Q_0$ and coupling $\beta$ for the $TM_{010}$ mode of the cavity is paramount for determining the expected power of the axions-conversion signal and its relation to the background noise. The $Q_0$ depends on the geometry of the cavity, requiring it to be measured after each step of the tuning process. To ensure accuracy and stability of the $Q_0$, which may be affected by mechanical rods stability and the electrical contact between rods and cavity walls, the measurement will be performed before and after each tuning step. A Vector Network Analyser (VNA) will be used to generate an RF sweep around the resonance frequency injected through a first ancillary RF line into the cavity through a weakly coupled calibration port. The $Q_0$ value will be deduced from the analysis of the transmission and reflection response.  The latter will be measured by a signal sent, from a second ancillary RF line, into the readout port through a directional coupler with coupling less than $-10$~dB. 
Gain calibration will be possible by combining the transmission coefficients between the ancillary and output lines~\cite{Alesini:2022lnp} or by connecting with a switch a 50$\Omega$ resistor to the amplifier, and varying its temperature with a heater. Additionally, the stability of the gain of the amplifiers can be checked by injecting a small known signal into the input port of the amplification chain. The measurement of the noise induced by room temperature electronics is less critical. One way to examine the external noise is to detach the cryogenic amplifier from the data collection system and attach it to a suitable termination resistor. The excess power detected in the acquired spectrum is a result of external interference. The speed at which this test can be performed depends on whether noise measurement is possible while the rod is in motion, the presence of external noise on the apparatus (which should be minimized in any case), and the analysis procedure.

Since the whole data acquisition will last for a rather long time (order of few years) it is important to guarantee periodically the stability of the reference clock with respect to an independent clock lock based on GPS. All the others working parameters (temperatures, pressures, magnetic field, etc.) are constantly measured and added to the data acquired.

We recall the integration time expressed in Eq.~\eqref{eq:integrationtime}. As stated in Sec.~\ref{section:III}, to have the sensitivity required for the measurement an integration time of 5 minutes for each frequency step with the large cavity and 10 minutes for the small cavity is foreseen. According to the number of frequency steps shown in Table~\ref{tab:tuning} for each cavity, this defines the total running time of about two years of data taking.  As described before, the power spectrum acquisition is done at the end of the AF stage with a commercial high-speed FFT. Assuming an analysis spectrum of 5 cavity widths, about 5\,kHz, each spectrum will be subdivided in 500 bins of 10 Hz each, a factor 10 less than the virialized-axion width. With a typical acquisition time of 10 ms per power spectrum, about 30000 spectra for the large cavity and 60000 spectra for the small cavity, will be acquired for each frequency step. The average of these spectra, for each step, will define the ``base spectrum'' for the considered frequency. 

Starting from the base spectra several steps are applied before the extraction of the final confidence limit as a function of the axion mass:
\begin{itemize}
\item definition of bad samples and problematic bins
\item normalization of the spectra
\item combination of spectra of residuals centered at different frequencies
\item test of the null hypothesis
\item re-scan of frequency regions that fail the null-hypothesis test
\item determination of the exclusion limit for frequency range that pass the null hypothesis test
\end{itemize}

The acquired spectra may be rejected for a variety of reasons, including fluctuations in resonance frequency, temperature changes, anomalous gain in amplifiers, and transient problems. The experiment slow-control system will be used for detecting and fixing long-standing issues, but will not guarantee that all data will be unaffected by temporary issues. Local and temporary issues such as electromagnetic interference may also lead to some bins of an acquired spectrum being affected. For all these reasons, it will be important to define a strategy to identify and mitigate the contribution of these effects. 


A standard approach to extract power spectrum residuals is to use a digital Savitzky-Golay (SG) filter. This filter is a polynomial generalization of a moving average which utilizes two parameters, W and d. It works by fitting a least-squares d-degree polynomial in a W-wide window for each value, x0, of the spectrum. This method is equivalent to a filter with a flat passband and mediocre stop-band attenuation~\cite{golay}. This works only if the cavity width is much larger than the intrinsic axion width, while for FLASH frequencies we expect large cavity quality-factors, about $5\times 10^5$. We will instead follow the approach described in Ref.~\cite{DiVora:2023dzs}. The power spectrum will be first described only by a ratio of first order polynomials or by means of a cavity equivalent-circuit. A null-hypothesis test will be performed at this stage on the $\chi^2$ of the fit. Runs failing this test will be subject to rescan to collect more statistics at specific frequencies, otherwise the fit residuals will be calculated. Given the short duration of data-taking run at different frequencies, 5 or 10 minutes, we do not expect large variations in the operational parameters during the run. The grand spectrum will be built by statistically combining the residuals from different runs in larger frequency-bins of about 100 Hz, corresponding to a virialized axion-width. Frequency bins with fluctuations above 3$\sigma$ will be subject to re-scan. This corresponds to an expected fraction of the frequency bins of about 0.135\% or to a 6\% re-scan probability of the run. In case the null-hypothesis test is passed, an upper limit will be calculated. The power spectrum will be now described as the sum of our cavity model and the axion signal, described by the standard halo model for dark matter. The coupling $g_{a\gamma\gamma}(m_a)$ will be determined by a fit to each power spectrum for different of values of the axion mass $m_a$. The couplings from different spectra fits are combined statistically according to the errors determined from the minimization procedure. Contrary to the S-G method, this fit procedure guarantees an efficiency close to one for the signal extraction.

The large amount of spectra needed to detect a fluctuation larger than the baseline at a specific confidence level necessitates the development of a strategy for data storage and processing. About 150000 frequency steps are foreseen to cover the whole frequency range accessible to FLASH. In each step an average of 45000 spectra will be collected, resulting in a total of about $10^{10}$ single spectra. The size of a single spectrum is determined by the digitizer. For a spectrum of 5\,kHz total width a digitization factor of 10 is necessary in order to identify axions conversion structures. Based on this, a single spectrum would consist of 8 kb assuming that the ADC is 16 bits. 
The total amount of data will be more than 10 TB. In addition, for each step, ancillary data will be collected. 
For the FFT calculation there are several solutions available beside a commercial one. One alternative is using a fast dedicated digitizer, as it would provide more information on the spectrum and enable FFT and filters to be applied at software level. An online pre-processing during data acquisition, utilizing the RAM of the computer as a temporary buffer, will compute, the average of the single spectra to obtain the baseline spectrum, needed to trigger the re-scan procedure, and it will include other steps of the proposed analysis, such as an algorithm for identifying problematic bins.  If the time required to transition between two steps of the procedure is enough to accommodate the online processing, then this should be feasible. If this is not the case, the process can be parallelized so that the processing of step N-1 can be completed while data is acquired for step N. For more complex processing, such as a Fourier Transform in software, the use of a Graphics Processing Unit (GPU) may be beneficial in order to enable fast, real-time processing without putting back pressure on the digitizer. Another option is to use an FPGA for digitization and online FFT calculation.

\subsection{High-frequency gravitational waves}

Signals emerging from the conversion of GWs from compact mergers and other coherent sources are expected to have a limited duration in a resonant mode, which means that the integration time for the signal in a resonant mode could not be increased above $\tau$, as defined in Sec.~\ref{ch:sec_GW}. Simultaneous data acquisition from different solenoidal normal modes of the cavity, with independent antennas and amplification circuits, would be of great importance for GW detection.

Due to the spin-2 nature of GWs, different normal modes of the cavity have different couplings to the direction of the incoming wave~\cite{Berlin:2022hfx}, so that multiple normal modes are required to improve the directional sensitivity pattern of the detector, while also allowing some constraints on the location of the source in the sky even with a single cavity~\footnote{Particularly interesting for directionality would be the use of a cubic cavity~\cite{benito2023}}. Most importantly, collecting data from multiple normal modes with different frequencies would also help improving the sensitivity and reach of the detector.

In fact, suppose the contribution to the power spectrum of the signal in different normal modes to be about equal and the thermal noise in different amplification circuits to be uncorrelated. Under these premises, the SNR of the detector is expected to scale at least as $\sim\sqrt{M}$ where $M$ is the number of measured modes that effectively couple with the gravitational wave signal. In realistic situations, mode mixing and lower coupling of some modes, especially ones of higher frequencies, might dampen this scaling. In principle, for signals with $N_{\rm cyc}\gtrsim 1$ a study in terms of Fourier spectrum, wavelets or time series could be implemented~\eqref{eq:strainberlinmatched}. 
A jump in sensitivity of several orders of magnitude will require to exploit the correlations of signals from many modes or cavities, as discussed in Ref.~\cite{Aggarwal:2020olq} and~\cite{PRXQuantum.3.030333}.
The correlations would have to be compared with a model and the likelihood maximized in order to estimate the physical parameters of the source. For related in-depth discussions see Refs.~\cite{Allen:2005fk, Maggiore:2007ulw, Moore:2014lga}. Note, that the GWs from mergers can couple with different normal modes at different times as the frequency evolves, which means that time discrepancies should be taken into account when exploiting the correlations for sufficiently long signals. The time between the detectable signals in the bandwidth of two different normal modes with $\nu_1<\nu_2$ could be estimated as~\cite{Maggiore:2007ulw}
\begin{equation}
    t_{\rm tr}\simeq 1.02\times10^{-8}{\rm\,s}\, \bigg( \frac{\mathcal{M}_{\rm c}}{10^{-5}\,M_{\odot}}\bigg)^{-\frac{5}{3}} \bigg(\frac{\nu_{1}}{200\text{ MHz}} \bigg)^{-\frac{8}{3}} \bigg[1 - \bigg(\frac{\nu_{1}}{\nu_{2}} \bigg)^{\frac{8}{3}} \bigg]\,
\end{equation}
where $\mathcal{M}_{\rm c}$ is the chirp mass. The correlation method could also be applied to the search for stochastic signals, although the expected characteristic strain from a predicted stochastic GWs background would be orders of magnitude lower than the sensitivity of FLASH.

The small power of the induced signal in the cavity and the resulting low SNR make the noise modeling and auxiliary data collection for detector monitoring extremely valuable for either a time series or spectrum analysis or in the correlations method. Multiplexing of more cavities or correlations with other independent detectors could also be implemented in the future to improve both SNR and source localization.

\section{Conclusions}
\label{sec:conclusions}

We have presented the activity undergoing at the INFN Frascati National Laboratories (INFN-LNF) regarding the proposed search for new physics with the FINUDA magnet for Light Axion SearcH or FLASH. The experiment will be able to probe the existence of dark matter composed by axions predicted by KSVZ and, with an upgraded cryogenic system, DFSZ models in the galactic halo within the mass range $m_a = (0.49-1.49)\,\mu{\rm eV}$. Together with the proposed BabyIAXO experiment the unexplored mass region up to 2$\mu$eV will be probed in the next decade. Other models for bosonic fields making up the cosmic dark matter such as an axion-like particle, scalar field or hidden photon can also be tested, as well as models of chameleons, which are not necessarily the dark matter, through the ``afterglow'' effect. Finally, the haloscope can be used to search for high-frequency gravitational waves of astrophysical origin. If the HFGW signal is due to coalescing PBHs, the experiment can probe the existence of mergers in the mass range $M_{\rm PBH} \lesssim 10^{-7}\,M_\odot$, although the expected strain from a single event in the Galaxy is too faint to be observed. We have presented the forecast reach of the FLASH experiment for all of the models above, along with a detailed explanation of the analysis involved. Forecasts have been presented for a detector with and without an improved cryogenic system able to reach temperatures down to $T_{\rm sys} = 100\,$mK. We stress here that in the latter case, the sensitivity of the experiment, even if limited by thermal noise, is still enough to probe KSVZ axions in the 1~$\mu$eV region, a result obtainable just by recycling the FINUDA magnet and its cryogenic plant. 

\appendix

\section{Computation of the couplings for the cavity modes TE$_{011}$ and TE$_{111}$}
\label{sec:coupling}

We compute the coupling in Eq.~\eqref{eq:coupling} for the cavity mode TE$_{011}$, whose EM field has components:
\begin{eqnarray}
	B_z &=& B_0\,J_0(v_0 \frac{r}{R})\,\sin(\pi \frac{z}{L})\,,\\
	B_r &=& -B_0\,\frac{\pi}{v_0}\frac{R}{L}\,J_1(v_0 \frac{r}{R})\,\cos(\pi \frac{z}{L})\,,\\
	B_\theta &=& 0\,,\\
	E_z &=& 0\,,\\
	E_r &=& 0\,,\\
	E_\theta &=& -iB_0\, \omega_0\,J_1(v_0 \frac{r}{R})\,\sin\left(\pi \frac{z}{L}\right)\,,
\end{eqnarray}
where $v_0 = 3.832$ is the first root of the Bessel function $J_1(x)$ and $\omega_0 = \left(1 + (\pi/v_0)^2(R/L)^2\right)^{1/2}$, where $R$ and $L$ are the cavity radius and length, respectively. We decompose the phase as $k\cdot x = k_r r\cos\theta + k_z z$, where $\theta \in [0,2\pi]$ is the angle between the radial direction and the projection of the momentum on the plane perpendicular to the cavity's axis, and we assume $k_z \approx k_r$. With this decomposition, the integral in Eq.~\eqref{eq:coupling} reads
\begin{eqnarray}
    \label{eq:coupling_011}
    C_{011} &=& \frac{1}{\pi^2\,L^2 R^2}\,\frac{\left|\int {\rm d}z \int {\rm d}\theta \int {\rm d}r r \,e^{i{\bf k}\cdot{\bf x}}\,J_0(v_0r/R)\,\sin(\pi z/L) \right|^2}{\int {\rm d}r r \, \left[J_0^2(v_0r/R) + \frac{\pi^2}{v_0^2}\frac{R^2}{L^2}\,J_1^2(v_0r/R)\right]} =\nonumber\\
    &=& \frac{\left|\frac{1 + e^{ik_z L}}{\pi^2-k_z^2L^2} \int_0^{2\pi} {\rm d}\theta \int_0^1 {\rm d}x x \,e^{ik_rR x\cos\theta}\,J_0(v_0 x) \right|^2}{\int_0^1 {\rm d}x x \, \left[J_0^2(v_0 x) + \frac{\pi^2}{v_0^2}\frac{R^2}{L^2}\,J_1^2(v_0 x)\right]}\,.
\end{eqnarray}
For the numerical estimate, we consider the large cavity setup with $R = 1050\,$mm and $L = 1200\,$mm. In the case of the chameleon for which the momentum is approximately the inverse Compton length $k_z^{-1} \approx k_r^{-1} \approx 0.2\,$m, the expression above gives $C_{011} \approx 0.005$. For bosonic DM with a velocity $v = 200\,$km/s, the momentum scale is set by the inverse de Broglie wavelength so that $k_z \approx mv \approx (200{\rm\,m})^{-1}$. In this case, the limits $k_z L \ll 1$ and $k_r R \ll 1$ lead to
\begin{equation}
    \label{eq:coupling_011approx}
    C_{011} = (mRv)^4 \frac{\left|\frac{1}{\pi} \int_0^1 {\rm d}x x^3\,J_0(v_0 x) \right|^2}{\int_0^1 {\rm d}x x \, \left[J_0^2(v_0 x) + \frac{\pi^2}{v_0^2}\frac{R^2}{L^2}\,J_1^2(v_0 x)\right]} \approx 0.003\,(mRv)^4\,,
\end{equation}
which gives $C_{011} \approx 5\times10^{-13}$ for the large cavity setup.

We repeat the computation for the cavity mode TE$_{111}$, whose EM field has components:
\begin{eqnarray}
	B_z &=& B_0\,J_1(v_1 \frac{r}{R})\,\sin\left(\pi \frac{z}{L}\right)\,\cos\theta\,,\\
	B_r &=& B_0\,\frac{\pi}{2v_1}\frac{R}{L}\,\left[J_0(v_1 \frac{r}{R}) - J_2(v_1 \frac{r}{R})\right]\,\cos\left(\pi \frac{z}{L}\right)\,\cos\theta\,,\\
	B_\theta &=& -B_0\,\frac{\pi}{v_1^2}\frac{R^2}{L\,r}\,J_1(v_1 \frac{r}{R})\,\cos\left(\pi \frac{z}{L}\right)\,\sin\theta\,,\\
	E_z &=& 0\,,\\
	E_r &=& iB_0\,\frac{\omega_1}{v_1}\frac{R}{r}\,J_1(v_1 \frac{r}{R})\,\sin\left(\pi \frac{z}{L}\right)\,\sin\theta\,,\\
	E_\theta &=& \frac{i}{2}\,B_0\,\omega_1\,\left[J_0(v_1 \frac{r}{R}) - J_2(v_1 \frac{r}{R})\right]\,\sin\left(\pi \frac{z}{L}\right)\,\cos\theta\,,
\end{eqnarray}
where $v_1 \approx 1.8412$ and $\omega_1 = \left(1 + (\pi/v_1)^2(R/L)^2\right)^{1/2}$. The integral in Eq.~\eqref{eq:coupling} for the mode TE$_{111}$ reads
\begin{equation}
    \label{eq:coupling_111}
    C_{111} = \frac{\left|\frac{1 + e^{ik_z L}}{\pi^2-k_z^2L^2} \int_0^{2\pi} {\rm d}\theta \int_0^1 {\rm d}x x \,e^{ik_rR x\cos\theta}\,J_1(v_1 x)\cos\theta \right|^2}{\frac{1}{2}\int_0^1 {\rm d}x x \, \left[ \left[1+\left(\frac{\pi R}{v_1^2Lx}\right)^2\right] J_1^2(v_1 x) + \left(\frac{\pi R}{2v_1L}\right)^2\,\left[J_0(v_1 x)-J_2(v_1 x)\right]^2 \right]}\,,
\end{equation}
which gives $C_{111} \approx 3\times 10^{-7}$ for the case of the large volume cavity, consistently with other findings~\cite{Flambaum:2022zuq}.

\section*{Acknowledgements}
The work of E.N.~was supported  by the Estonian Research Council grant PRG1884, by the INFN ``Iniziativa Specifica'' Theoretical Astroparticle Physics (TAsP-LNF). C.G., M.G., E.N., L.V., and M.Z.\ acknowledge the Galileo Galilei Institute for Theoretical Physics in Florence for hospitality during the completion of this work. L.V.\ and M.Z.\ acknowledge hospitality by the INFN Frascati National Laboratories, as well as support by the NSFC through the grant No.\ 12350610240. This publication is based upon work from the COST Actions ``COSMIC WISPers'' (CA21106) and ``Addressing observational tensions in cosmology with systematics and fundamental physics (CosmoVerse)'' (CA21136), both supported by COST (European Cooperation in Science and Technology).  B.D.\  acknowledges support by the European Research Council under grant ERC-2018-StG-802836 (AxScale project). The Authors acknowledge the the LNF services and divisions for the technical support received for the ongoing test of the FINUDA magnet in particlar Marco Beatrici, Daniele Di Bari, Giuseppe Ceccarelli, Franco Iungo, Ruggero Ricci, Luigi Pellegrino, Luca Piersanti, Sandro Gallo, Andrea Ghigo, Bruno Buonomo, Sandro Vescovi, Ugo Rotundo and Sergio Cantarella.

\bibliographystyle{elsarticle-num}
\bibliography{FLASH.bib}

\end{document}